\documentclass[12pt,pra,nofootinbib,preprint,pdftex]{revtex4}
\usepackage{amsmath,amssymb}
\usepackage[dvips]{graphicx,color}
\usepackage{epstopdf}
\usepackage[pdftex,colorlinks]{hyperref}
\newcommand{\transpose}{\!\scriptscriptstyle\mathrm T}
\newcommand{\bolden}[1]{\mbox{\boldmath$#1$}}
\numberwithin{equation}{section}
\begin{document}

\title{Solutions to Master Equations of Quantum Brownian Motion in a
General Environment with External Force}
\author{C. H. Fleming, B. L. Hu}
\affiliation{Joint Quantum Institute and Department of Physics, University
of Maryland, College Park, Maryland 20742}
\author{Albert Roura}
\affiliation{Theoretical Division, T-8, Los Alamos National Laboratory,
M.S.~B285, Los Alamos, NM 87545}
\date{\today}

\begin{abstract}
We revisit the model of a system made up of a Brownian quantum
oscillator under the influence of an external classical force and
linearly coupled to an environment made up of many quantum oscillators
at zero or finite temperature.  We show that the HPZ master equation
for the reduced density matrix derived earlier [B.L.~Hu, J.P.~Paz,
Y.~Zhang, Phys.\ Rev.~D \textbf{45}, 2843 (1992)] with coefficients
obtained from solutions of integro-differential equations can assume
closed functional forms for a fairly general class of spectral
densities of the environment at arbitrary temperature and coupling
strength.  As an illustration of these new results we solve the
corresponding master equation and calculate, among other physical
quantities, the uncertainty function whose late time behavior can be
obtained fully.  This produces a formula for investigating the
standard quantum limit which is central to addressing many theoretical
issues in macroscopic quantum phenomena and experimental concerns
related to low temperature precision measurements.  We find that any
initial state always settles down to a Gaussian density matrix whose
covariance is determined by the thermal reservoir and whose mean is
determined by the external force.  For more general spectra we show
that the solution of the master equation can be reduced to solving for
the motion of a classical parametric oscillator with parametric
frequency determined by the \emph{unsolved for} master equation
coefficients.  States in these systems experience evolution that is
parametrically similar to the simpler evolution explicitly determined
for in the case of Laurent-series spectra.
\end{abstract}

\maketitle

\tableofcontents

\section{Introduction}

\subsection{Historical Context and New Results}

In this paper we continue the lineage of work on quantum Brownian
motion via the influence functional path-integral method of Feynman
and Vernon\cite{feynman} used by Caldeira and Leggett\cite{leggett} to
derive a master equation for a high-temperature ohmic environment,
which corresponds to the Markovian regime.  Following this, Caldeira,
Cerdeira and Ramaswamy (CCR)\cite{caldeira} derived the Markovian
master equation for the system with weak coupling to an ohmic bath,
that was claimed to be valid at arbitrary temperature
(Sec.~\ref{sec-caldeira}).  At the same time Unruh and
Zurek\cite{unruh} derived a more complete and general master equation
that incorporated a colored noise at finite temperature.  Finally Hu,
Paz and Zhang (HPZ) \cite{QBM1} derived an exact master equation for a
general environment (arbitrary temperature and spectral density) which
preserves the positive definiteness of the reduced density matrix, an
important property missed out in many earlier derivations.

For many calculations, physicists often invoke Markovian master
equations as they are easier to solve owing to their having compact,
functional representations.  But one runs into trouble if one is
interested in low temperatures, short times or non-ohmic baths, which
likely fall under the non-Markovian regime (see Ref.~\cite{ShiHu04}
for a discussion).  The HPZ equation is capable of dealing with the
full range of parameters for a general environment but its form is
somewhat involved.  For example the coefficients are represented by
solutions to integro-differential equations and multiple integrals
over Green's functions.  In this paper we present the correct
(\emph{cf.} for instance CCR) late-time non-Markovian master equation
in compact functional representation for arbitrary temperature and
coupling strength (Sec.~\ref{sec:latediff}).  Additionally we obtain
the coefficients for a large class of spectral densities and their
time dependent form at times before the oscillator and reservoir have
equilibrated.

Solutions to the HPZ equation have been attempted before
\cite{solnHPZ} and the role of the master equation coefficients
(renormalized frequency, dissipation, diffusion, and anomalous
diffusion coefficients) has been discussed in the past.  Here we
present complete solutions of the master equation.  We consider a
quantum oscillator under the influence of an external classical force
that is linearly coupled to a thermal reservoir of quantum
oscillators.  The particle begins decoupled from the reservoir and
uncorrelated with it, with a short switch-on time $\tau$ for the
coupling.  The master equation coefficients are derived for spectral
functions which are Laurent series in the frequency with ultraviolet
(UV) cut-off $\Lambda$ much larger than the system frequencies, and
infrared (IR) cut-off $\lambda$ much smaller than the system
frequencies. These coefficients are calculated in detail for
arbitrary temperature and arbitrary time after the switch-on time.
The master equation is solved explicitly for this class of spectra,
and the general solution for arbitrary coefficients is reduced to
solving the classical equation of motion of a parametric oscillator.

It is clearly shown how each coefficient enters into the solution and
how the state evolves in time.  The anomalous diffusion coefficient is
actually an ``anti-diffusion'' term that keeps the position
uncertainty finite even when the UV cut-off tends to infinity (in
contrast with the momentum uncertainty).  Having the solution gives us
a plethora of information such as the late time thermal covariance
[Eq.~\eqref{eq:ltsig}] and uncertainty function (see
Sec.~\ref{sec:uncertainty}).  These results generalize the work of
Anastopoulos and Halliwell \cite{halliwell}, who already found the
late time state to be a Gaussian and placed lower bounds on the
uncertainty function which we discuss in Sec.~\ref{sec:uncertainty}.
This also crystallizes the work of Hu and Zhang \cite{HZ} on the
generalized uncertainty function for Gaussian states.

Finally, we derive the master equation that includes the influence of
a classical external force and solve the master equation in that case
as well.

\subsection{Systematic Overview}

The paper is organized as follows.  We begin our derivation of the
master equation in Sec.~\ref{sec:master-ohmic} with a quantum
oscillator linearly coupled to a thermal reservoir of quantum
oscillators.  The spectral density here is assumed to be ohmic with
high-frequency cut-off $\Lambda$.  The particle begins both
\emph{decoupled} from and \emph{uncorrelated} with the reservoir. The
coupling is turned off at the initial time $t=0$ and is switched on
within a short time scale $\tau$. At moderate times, $t \gg \tau$ (but
not necessarily $t>\Omega_r^{-1}, \gamma_0^{-1}$), we have obtained
the master equation coefficients as an expansion from zero temperature
(Sec.~\ref{sec:master-low}), as an expansion from high temperature
(Sec.~\ref{sec:master-high}), and in closed form as an approximation
that is exact at both zero and extreme temperatures
(Sec.~\ref{sec:master-approx}).  Exact closed form solutions are
possible in terms of various special functions, but they do not
straightforwardly reveal the behavior seen in our expansions and
approximations.  Most importantly, we have closed form solutions for
the master equation coefficients at late times, $t\gg \gamma_0^{-1}$,
and arbitrary temperature (Sec.~\ref{sec:master-late}).

In Sec.~\ref{sec:master-laurent} we extend these results first to
analytic spectral functions (Sec.~\ref{sec:master-power}) and then to
Laurent-series spectra (Sec.~\ref{sec:master-laurent2}).  With the
inclusion of subohmic terms we must introduce a low frequency cut-off
$\lambda$.  We take both the UV and IR cut-offs to be very large and
very small respectively and we only consider spectra that give
contributions to the master equation coefficients no more divergent
than in the ohmic case (which has logarithmic dependence on the UV
cut-off).
This limits our study to equations of motion for the system
trajectories with the same form as in the ohmic case and whose
contributions to the master equation coefficients are not markedly
different either.  Spectral densities of arbitrary powers have been
studied numerically in the past, but we do not consider them as they
correspond to classical paths that involve fractional calculus
(\emph{i.e.}, integro-differential equations) and would be difficult
to solve analytically.

In Sec.~\ref{sec:wigner} we solve the master equation for this class
of spectra.  It is seen that the initial solution undergoes damped
oscillations while evolving into a Gaussian state of thermal
equilibrium.  All the cumulants of the Wigner distribution are easily
determined as functions of time (Sec.~\ref{sec:wigner-sol}).  In
particular we can provide the exact late-time uncertainty function
(Sec.~\ref {sec:uncertainty}).  In Sec.~\ref{sec:wigner-gen} we extend
this solution to more general spectra, provided that one knows the
form of the master equation coefficients.  The solution to this master
equation is reduced to solving for the motion of a classical
parametric oscillator.  It is seen that simple damped oscillations may
now be parametric damped oscillations, otherwise the form of the
solution is the same.

Finally in Sec.~\ref{sec:force} we extend the method of Calzetta,
Roura, and Verdaguer \cite{roura} to include the influence of a
classical force acting upon the oscillator.  The master equation turns
out to be what one would naively guess.  We solve this master equation
in Sec.~\ref{sec:force-sol} and it is seen that the external force
drives the mean around while the evolution of the variance remains
unchanged and determined by the reservoir.

In the last section we conclude with a list of our findings and
suggestions for their applications.

\section{Master Equation Coefficients for an Ohmic Spectrum }
\label{sec:master-ohmic}

\subsection{Ohmic Spectrum Master Equations}

The Lagrangian of a system consisting of a quantum Brownian
oscillator with mass $M$, natural frequency $\Omega$ and coordinate
$x$ coupled with coupling constants $c_n$ to an environment at
temperature $T$ made up of $n$ oscillators with mass $m_n$, natural
frequency $\omega_n$ and coordinates $x_n$ is given by
\begin{equation}
L = \frac{1}{2} M \left( \dot{x}^2 - \Omega^2 x^2 \right) + \sum_n \frac{1}{2} m_n \left( \dot
{x}_n^2 - \omega_n^2 x_n^2 \right) - \theta_\tau(t) \sum_n c_n x x_n
\label{eq:switch},
\end{equation}
where $\theta_\tau(t) \approx e^{-\left(\frac{\tau}{t}\right)^2}$ is
a switch-on function with a very short characteristic time-scale
$\tau$.  In this section we work with an ohmic bath so that the
spectral density function is given by
\begin{equation}
\label{eq:spectrum} I(\omega) = \frac{2}{\pi} \gamma_0 M \omega \qquad (0 < \omega <
\Lambda) ,
\end{equation}
where $\Lambda$ is a high-frequency cut-off of the bath and $\gamma_0$ is the
dissipation frequency.
We will be using this hard cut-off for all our calculations.
Gaussian and exponential decay cut-off functions, of the form $e^{-(\omega /
\Lambda)^2}$ and $e^{-\omega / \Lambda}$ respectively, are also common
choices.

The system and environment begin decoupled at $t=0$, but are fully
engaged in a time $ \tau$ which is assumed to be very short but
non-zero.  This time-dependent coupling is easily dealt with following
the work of Hu and Matacz\cite {QBM3} on the time-dependent problems
where all parameters of the system and bath oscillators and their
couplings are allowed to be time-dependent.  When only the
system-environment couplings are time-dependent, as in our case, the
dissipation and noise kernels are given respectively by
\begin{eqnarray}
\label{eq:dissipation} \mu(t,s) & = & 2 \gamma_0 M \frac{d}{dt} \delta_\Lambda(t-s) \mbox
{~} \theta_\tau(t) \theta_\tau(s) , \\
\label{eq:noise} \nu(t,s) & = & \int_0^\infty d\omega \cos{[\omega(t-s)]} \coth{\left( \frac
{\omega}{2 T} \right)} I(\omega) \mbox{~} \theta_\tau(t) \theta_\tau(s) ,
\end{eqnarray}
where
\begin{equation}
\delta_\Lambda(t) = \frac{1}{\pi} \int_0^\Lambda d\omega \cos{(\omega t)}
= \frac{\sin (\Lambda t)}{\pi t}
\label{eq:delta1},
\end{equation}
and it tends to a Dirac delta distribution in the limit $\Lambda
\rightarrow \infty$.  Had we chosen a Gaussian high-frequency cut-off
function instead of a hard cut-off, we would have found the
dissipation kernel \eqref{eq:dissipation} to be the derivative of a
Gaussian distribution that corresponds to an alternative
representation of the delta distribution in the infinite cut-off
limit.  On the other hand, the integral for the noise kernel
\eqref{eq:noise} is more complicated, even without the cut-off.
Although it can also be expressed in terms of distributions, for the
purposes of this paper it is best left to the last.

\subsubsection{Initial Time Divergences}

In HPZ\cite{QBM1} the system-environment coupling was taken to be
constant in time, yet the system and environment were initially
uncorrelated. This gives rise to initial ``jolts'' in the normal
diffusion coefficient of the master equation with a characteristic
time-scale of order $\Lambda^{-1}$ and an amplitude proportional to
$\Lambda$,
which diverges in the limit $\Lambda \rightarrow \infty$. It also
gives rise to a delta-like term at the initial time in the equation of
motion:
\begin{equation}
\ddot{u}(t) + 2 \gamma_0 \dot{u}(t) + \Omega_r^2 u(t) = -4 \gamma_0 \delta_\Lambda(t)
u(0) .
\label{eq:eom1}
\end{equation}
In the limit of infinite cut-off this sudden frequency change gives an
initial kick to the homogeneous solutions of the equation of motion,
which can be seen as follows:
\begin{align}
\dot{u}         &= v                                    
&   \Delta u = \lim_{\epsilon \to 0}\int_0^ \epsilon dt \dot{u} &= 0\\
\dot{v}         &= -2 \gamma_0(t) v - \Omega_r^2(t) u   
&   \Delta v = \lim_{\epsilon \to 0}\int_0^\epsilon dt \dot{v} &= -2\gamma_0 u(0)\\
\Omega_r^2(t)   &= \Omega_r^2 + 4 \gamma_0 \delta(t) &&
\end{align}
Thus, the classical paths experience a finite velocity change within
an infinitesimal time.  Following the approach in Ref.~\cite{roura}
one can easily see that this kick translates into a distortion of the
Wigner distribution from the bare initial state to a shifted one
\begin{eqnarray}
W_\mathrm{bare}(x,p) &\to& W_\mathrm{ren}(x,p) = W_\mathrm{bare} (x,p+2 M \gamma_0 x)
\label{eq:phase_trans1} .
\end{eqnarray}
One expects that this result, which corresponds to replacing
$\delta_\Lambda(t)$ with a Dirac delta in Eq.~(\ref{eq:eom1}), is a
good approximation whenever all the relevant time-scales
($\Omega_r^{-1}$, $\gamma_0$ and $t$) are much larger than $\Lambda^
{-1}$.

The physical origin of the jolts in the coefficients of the master
equation as well as other initial time divergences (such as the
divergent contributions to correlation functions of system observables
that are due to divergent boundary terms at the initial time (see
Appendix~D in Ref.~\cite{HRV}) can be understood as follows. In
general when a system couples to an environment with an infinite
number of modes, well-behaved states exhibit correlations with
arbitrarily high-frequency modes. In contrast, states that are
uncorrelated for sufficiently high frequencies (and hence completely
factorizable states in particular) are pathological. For instance,
they have infinite energy and their Hilbert space is even unitarily
inequivalent to the space of physical states, spanned by the basis of
energy eigenvectors of the whole system Hamiltonian including the
system-environment interaction. (Of course for a finite UV cut-off
there are no divergences, but the potentially divergent terms are very
sensitive to changes in the value of the cut-off.) Physically
acceptable initial states that correspond to the thermal equilibrium
state for the whole system can be obtained using Euclidean path
integrals \cite{grabert}. However, the instantaneous preparation
functions employed in Ref.~\cite{grabert} to produce other states in
addition to the thermal equilibrium state still give rise to initial
divergences, as explained in Ref.~ \cite{romero}. In order to obtain
finite results, one needs to prepare the new initial state within a
non-vanishing time \cite{anglin}, which corresponds to a physically
more realistic situation. The alternative approach that we follow here
is to switch on the interaction smoothly within a time $\tau$ much
longer than $\Lambda^{-1}$ but shorter than any other relevant
time-scale of the problem. In this way the factorized initial state,
which is perfectly acceptable in the uncoupled case, becomes
adequately correlated with the arbitrarily high-frequency modes in a
regular fashion.

When adding the short time switch-on function \eqref{eq:switch} to the
spectral density to turn on the interaction gradually, the initial
jolt is no longer present in the results for the master equation
coefficients. Moreover, the equation of motion exhibits a smooth
transition between the decoupled and the coupled system without the
term proportional to $\delta_\Lambda(t)$ on the right-hand side:
\begin{eqnarray}
\ddot{u}(t) + \Omega^2 u(t) &=& 0,  \quad t \ll \tau \\
\ddot{u}(t) + 2 \gamma_0 \dot{u}(t) + \Omega_r^2 u(t) &=& 0,  \quad \tau \ll t
\label{eq:eom2}.
\end{eqnarray}
However, since the constant bare frequency $\Omega^2$ is of order
$\Lambda$ so as to cancel out the divergent contribution that arises
when integrating out the environment and give a finite value of the
renormalized frequency at late times (larger than the switch-on time),
the frequency will change significantly in a short period of time of
order $\tau$.  As long as $\tau$ is much shorter than all the other
relevant time-scales in the problem (except for $\Lambda^{-1}$), the
dynamics for $t \gg \tau$ can be obtained by approximating the
time-dependent frequency by a delta function. More specifically, since
the renormalized frequency changes within a characteristic time-scale
of order $\tau$ and with an amplitude of order $\Lambda$, it can be
written as
\begin{equation}
\Omega_r^2 (t) = \Omega_r^2 - 2 c \Lambda \tau \, \delta_\tau(t) ,
\label{eq:ren_freq}
\end{equation}
where $\Omega_r^2$ is the asymptotic constant value for times much
larger than $\tau$, $c$ is a constant of order one, and
$\delta_\tau(t)$ is a function peaked at $t=0$ with amplitude of order
$\tau^{-1}$ and width of order $\tau$ that becomes a Dirac delta in
the limit $\tau \rightarrow 0$.  Therefore, for $t \gg \tau$ the
dynamics is governed by
\begin{equation}
\ddot{u}(t) + 2 \gamma_0 \dot{u}(t) + \Omega_r^2 u(t) = 2 c \, \Lambda \tau \, \delta (t) u (0).
\label{eq:eom3}
\end{equation}
Note that $\Lambda \tau \gg 1$ if the switch-on time is sufficiently
large to cure the initial divergences discussed above.

The term on the right-hand side of Eq.~(\ref{eq:eom3}) has the same
form as in Eq.~ (\ref{eq:eom1}). It will, therefore, give the same
kind of initial kick to the solutions of the equation and generate the
same kind of transformation of the reduced Wigner function:
\begin{equation}
W_\mathrm{bare}(x,p) \to W_\mathrm{ren}(x,p) = W_\mathrm{bare} (x,p - c \, \Lambda \tau M x)
\label{eq:phase_trans2} .
\end{equation}
This phase-space transformation has a Jacobian with determinant equal
to unity:
\begin{equation}
L = \left( \begin{array}{cc} 1 & 0 \\ -c \, \Lambda \tau M & 1 \end{array} \right)
\qquad  \det L =1 .
\label{eq:jacobian}
\end{equation}
Therefore, it is simple to calculate renormalized expectation values
in terms of bare expectation values and vice versa:
\begin{eqnarray}
\langle A(x,p) \rangle_\mathrm{ren} &=& \iint dx dp A(x,p) W_\mathrm{ren}(x,p) , \\
\langle A(x,p) \rangle_\mathrm{ren} &=& \langle A(x,p+c \, \Lambda \tau M x) \rangle_
\mathrm{bare} \, .
\end{eqnarray}
We can immediately see that the normalization, linear entropy (see
Sec.~\ref {sec:entropy}) and state overlap are all unchanged by the
kick.  We can also check that the Heisenberg uncertainty relation is
also preserved as follows.  First, we start with the covariance matrix
for $x$ and $p$ corresponding to the Wigner distribution
\begin{equation}
\bolden{\sigma} = \left( \begin{array}{cc} \sigma_{xx} & \sigma_{xp} \\ \sigma_{px} &
\sigma_{pp} \end{array} \right) ,
\label{eq:covariance}
\end{equation}
with $\sigma_{xx} = \langle xx \rangle_\mathrm{ren}$, $\sigma_{xp} =
\sigma_{px} = \langle xp \rangle_\mathrm{ren}$ and $\sigma_{pp} =
\sigma_{pp} = \langle pp \rangle_\mathrm{ren}$, and which transforms
in the following way under linear phase space transformations:
\begin{equation}
\bolden{\sigma} \to L^T \bolden{\sigma} L \, .
\end{equation}
Hence, from Eq.~(\ref{eq:jacobian}) we have
\begin{equation}
\det{\bolden{\sigma}}_{bare} = \det{\bolden{\sigma}}_{ren} \, .
\end{equation}
Finally, one takes into account that
\begin{equation}
(\det{\bolden{\sigma}}) \geq \frac{\hbar^2}{4}
\label{eq:uncertainty1},
\end{equation}
corresponds to the formulation in terms of the Wigner function of the
generalized Heisenberg uncertainty relation due to Schr\"odinger
\cite{robertson,trifonov}:
\begin{equation}
(\Delta x)^2 (\Delta p)^2 -  \left\langle \{\hat{x} - \langle \hat{x} \rangle,\hat{p}
-\langle \hat{p} \right\rangle\} \rangle^2 \geq \frac{\hbar^2}{4}
\label{eq:uncertainty2},
\end{equation}
where $\{\hat{A},\hat{B}\} \equiv \hat{A}\hat{B} + \hat{B}\hat{A}$.
In addition to being real and normalized, a phase space distribution
corresponds to the Wigner function of a physical state only when the
corresponding density matrix (undoing the Wigner transformation)
satisfies the condition $\rho^2 \leq \rho$. For Gaussian distributions
this is guaranteed if and only if Eq.~(\ref{eq:uncertainty1}) is
satisfied \cite {kruger}. Thus, the kick does not change whether a
given Gaussian distribution corresponds to a physical state. On the
other hand, for non-Gaussian distributions Eq.~(\ref
{eq:uncertainty1}) is no longer a sufficient condition for that, but
one can use the following argument. For a finite switch-on time, the
frequency change simply corresponds to unitary evolution associated
with a time-dependent Hamiltonian, which leaves the eigenvalues of the
density matrix non-negative.  By continuity, the limit of a very short
switch-on time cannot make these eigenvalues negative.

If one is interested in studying the evolution of a certain state of
the system properly correlated with the environment (such as the
global equilibrium states considered in Ref.~ \cite{grabert} or states
prepared from those in a finite time) one can always consider the
reduced Wigner function associated with that state and invert
Eq.~(\ref{eq:uncertainty1}) to obtain the corresponding initial Wigner
function before the interaction was switched on. (Note that for times
larger than the switch-on time the result is then essentially
equivalent to having introduced a time-dependent counterterm for the
frequency so that the renormalized frequency was constant in time.)

\subsubsection{Expressions for the Master Equation Coefficients}

The HPZ master equations for the reduced density matrix $\rho_r$ and
the reduced Wigner function are given respectively by
\begin{eqnarray}
\imath \frac{\partial}{\partial t} \hat{\rho}_r &=& \left[ \hat{H}_r, \hat{\rho}_r \right] +
\Gamma \left[\hat{x},\{\hat{p},\hat{\rho}_r\}\right]
+ \imath D_{xp} \left( \left[\hat{x},\left[\hat{p},\hat{\rho}_r\right]\right]
+ \left[\hat{p},\left[\hat{x},\hat{\rho}_r\right]\right] \right) - \imath
D_{pp} \left[\hat{x},\left[\hat{x},\hat{\rho}_r\right]\right] , \\
\frac{\partial}{\partial t} W_r &=& \{ H_r,W_r \} + 2\Gamma \frac{\partial}
{\partial p} p W_r -2 D_{xp} \frac{\partial^2}{\partial x \partial p} W_r + D_{pp} \frac
{\partial^2}{\partial p^2} W_r
\label{eq:wigner},
\end{eqnarray}
where the detailed form of the dissipation function $\Gamma(t)$ and
the diffusion functions $D_{xp}(t)$ and $D_{pp}(t)$ can be found in
\cite{QBM1}.  Following the derivation of Calzetta, Roura, and
Verdaguer \cite{roura} the coefficients in the HPZ master equation for
an ohmic bath with a high cut-off $ \Lambda$ and evaluated at times
larger than the switch-on time are given by\footnote{This method has
  the nice property that the master equation coefficients after the
  switch-on time are explicitly independent of any reasonable behavior
  during the initial switch-on.}:
\begin{eqnarray}
\Omega_\mathrm{ren}^2 &=& \Omega_r^2 \, , \\
\Gamma &=& \gamma_0 , \\
D_{xp} &=& -\frac{1}{2} \int_0^t ds \mbox{~} \nu(t,s) G_\mathrm{ret}(t,s)
\label{eq:Dxp1}, \\
D_{pp} &=& M \int_0^t ds \mbox{~} \nu(t,s) \frac{\partial}{\partial t} G_\mathrm{ret}(t,s)
\label{eq:Dpp1} , \\
G_\mathrm{ret} (t,s) &=& \frac{1}{M \tilde{\Omega}} \sin \tilde{\Omega}(t-s) \,
e^{-\gamma_0(t-s)}
\label{eq:green},
\end{eqnarray}
where $\tilde{\Omega} = \sqrt{\Omega_r^2 - \gamma_0^2}$ and
$G_\mathrm{ret} (t,s)$ is the retarded Green function associated with
the differential equation that results from multiplying
Eq.~(\ref{eq:eom2}) by $M$.
According to the discussion in the previous subsection, when
considering times much larger than the switch-on time, its effect on
the equation of motion can be approximated by a delta function, as
seen in Eq.~(\ref{eq:eom3}). Since $G_\mathrm{ret} (t,s)$ for $t \leq
0$, the contribution from the delta term on the right-hand side of
Eq.~(\ref{eq:eom3}) vanishes when solving for the retarded
propagator. Therefore, it is indeed independent of the initial kick
and given by Eq.~(\ref{eq:eom3}). There is still some dependence on
the switch-on function $\theta_\tau (s)$ in the noise kernel for $s <
\tau$, but for $t \gg \tau$ the contribution from $s < \tau$ to the
time integrals in Eqs.~(\ref{eq:Dxp1})-(\ref {eq:Dpp1}) is negligible
provided that the behavior during the switch-on time is sufficiently
regular, which is indeed the case. Similarly, the behavior of the
master equation coefficients during switch-on time, which we do not
calculate here, will be sufficiently regular so that its contribution
to the evolution of the reduced Wigner function (or density matrix) is
also negligible when considering times larger than $\tau$ and one only
needs to include the effect of the kick on the initial Wigner
function, given by Eq.~(\ref{eq:phase_trans2}).

Some clarification is in order here about the use of a finite
frequency cut-off $\Lambda$ for the environment spectrum. Having a
finite cut-off in the frequency integral for the noise kernel, given
by Eq.~(\ref{eq:noise}), is necessary to obtain a finite result for
the coefficients of the master equation because, as it will be seen
below, it gives rise to contributions to the diffusion coefficients
that become logarithmically divergent in the limit of large
$\Lambda$. In contrast, the use of a finite cut-off in the frequency
integral of the dissipation kernel is not required to obtain a finite
result. In fact, for computational convenience we are going to take
$\Lambda \to \infty$ in this case because then the dissipation kernel
becomes local and we just have to deal with an ordinary differential
equation rather than an integro-differential one. One may object that
treating the noise and dissipation kernels on a different footing
could lead to inconsistencies (\emph{e.g.}  the
fluctuation-dissipation relation will not hold exactly any more,
as pointed out by HPZ \cite{QBM1} with regard to the treatment of
this problem in Ref.~\cite{unruh}). However, our results are consistent
when properly understood. If one considered an expansion of the exact
master equation coefficients in terms $\Lambda$, one would have
leading terms of order $\log \Lambda$, terms of order one and
subdominant terms involving inverse powers of $\Lambda$. Our approach
gives the right result for the $\log \Lambda$ and order-one terms:
using the local approximation for the dissipation kernel only alters
the result for the subdominant terms. This can be seen as follows.
When integrating $\delta_\Lambda(t-s) $, given by
Eq.~(\ref{eq:delta1}), with a sufficiently well-behaved function
$f(s)$, one has
\begin{equation}
\int_{-\infty}^\infty ds \, \delta_\Lambda(t-s) f(s) = f(t) + f'(t) \Lambda^{-1} + O(\Lambda^
{-2}) .
\end{equation}
This condition will indeed be satisfied in our case since all the
other relevant time-scales in the problem are much larger than
$\Lambda^{-1}$ and all the functions that $\delta_ \Lambda(t-s)$ will
be integrated with are regular and extremely uniform over that scale.
Therefore, the discrepancy between using the Dirac delta function and
using $\delta_\Lambda(t-s)$ corresponds to terms of order
$\Lambda^{-1}$ or higher inverse powers, which give contributions to
the master equation coefficients of order $\Lambda^{-1} \log \Lambda$
or higher.

In order to compute the coefficients of the master equation, we first
perform the relatively simple time integrals in
Eqs.~(\ref{eq:Dxp1})-(\ref{eq:Dpp1}). On the other hand, the frequency
integral in the noise kernel, as given by Eq.~(\ref{eq:noise}), is
particularly difficult due to the $1 / \sinh({\omega}/{2T})$ in
$\coth({\omega}/{2T})$. It is this integral that we save for last.
Others\cite{grabert,unruh} have also seen the utility or at least the
simplicity of reducing the master equation to a collection of
one-dimensional frequency integrals.  We are able to obtain four sets
of new results: (a) exact results at late time for arbitrary
temperature, (b) asymptotic expansions for high and low temperatures,
(c) approximate general results for arbitrary temperature and all
times (larger than the switch-on time), (d) time-dependent diffusion
functions in closed form at high temperatures for all times (larger
than the switch-on time).

After computing the time integrals in Eqs.~(\ref{eq:Dxp1}) and
(\ref{eq:Dpp1}), the diffusion coefficients become
\begin{eqnarray}
D_{xp}(t) &=& +\frac{\gamma_0}{\pi}\left( \mbox{FI}_3 -  \Omega_r^2 \mbox{FI}_1 \right)\nonumber \\
&& - \frac{\gamma_0}{\pi} \cos(\tilde{\Omega} t) e^{-\gamma_0 t}\left( \mbox{FC}_3(t)
- \Omega_r^2 \mbox{FC}_1(t) + 2 \gamma_0 \mbox{FS}_2(t) \right) \nonumber  \\
&& + \frac{\gamma_0}{\pi \tilde{\Omega}} \sin(\tilde{\Omega} t) e^{-\gamma_0 t}
\left( \gamma_0 \left( \mbox{FC}_3(t) + \Omega_r^2 \mbox{FC}_1(t) \right) + \left( \tilde
{\Omega}^2 - \gamma_0^2 \right) \mbox{FS}_2(t) - \mbox{FS}_4(t) \right) ,\label{eq:Dxp2} \\
D_{pp}(t) &=& +\frac{4 M \gamma_0^2}{\pi} \mbox{FI}_3 \nonumber \\
&& - \frac{2 M \gamma_0}{\pi} \cos(\tilde{\Omega} t) e^{-\gamma_0 t}\left( 2 \gamma_0
\mbox{FC}_3(t) + \Omega_r^2 \mbox{FS}_2(t) - \mbox{FS}_4(t) \right) \nonumber \\
&& - \frac{2 M \gamma_0}{\pi \tilde{\Omega}} \sin(\tilde{\Omega}t) e^{-\gamma_0 t}\left( -
\Omega_r^4 \mbox{FC}_1(t) + \left(\tilde{\Omega}^2 - \gamma_0^2 \right)\mbox{FC}_3(t)
+ \gamma_0\left( \Omega_r^2 \mbox{FS}_2(t) + \mbox{FS}_4(t) \right) \right) ,
\nonumber\\
\label{eq:Dpp2}
\end{eqnarray}
in terms of a single family of frequency integrals $\mbox{FC}_N$,
$\mbox{FS}_N$, $ \mbox{FI}_N$:
\begin{eqnarray}
\mbox{FI}_N & = & \int_0^\Lambda \frac{\omega^N \coth\left( \frac{\omega}{2T} \right)}
{\left( \omega^2-\Omega_r^2 \right)^2 + 4 \gamma_0^2 \omega^2} d\omega
\label{eq:FI_N}, \\
\mbox{FC}_N(t) & = & \int_0^\Lambda \frac{\omega^N \cos(\omega t) \coth\left( \frac
{\omega}{2T} \right)}{\left( \omega^2-\Omega_r^2 \right)^2 + 4 \gamma_0^2 \omega^2}
d \omega \label{eq:FC_N}, \\
\mbox{FS}_N(t) & = & \int_0^\Lambda \frac{\omega^N \sin(\omega t) \coth\left( \frac
{\omega}{2T} \right)}{\left( \omega^2-\Omega_r^2 \right)^2 + 4 \gamma_0^2 \omega^2}
d \omega , \\
\mbox{FI}_N & = & \mbox{FC}_N(0) , \\
\mbox{FS}_N(t) & = & - \frac{d}{dt} \mbox{FC}_{N-1}(t) , \\
\mbox{FC}_N(t) & = & + \frac{d}{dt} \mbox{FS}_{N-1}(t) .
\end{eqnarray}
Fortunately the only integral that needs to be computed is
$\mbox{FC}_1(t)$. All other integrals can be generated from this one
integral.  In general the integrals cannot be solved directly,
therefore it is necessary to expand the hyperbolic cotangent into a
series of simpler functions.  This will be done with low and
high-temperature expansions.

When computing the time integrals that led to
Eqs.~(\ref{eq:Dxp2})-(\ref{eq:Dpp2}), we just took the switch-on
functions equal to one, so that sufficiently simple analytical results
could be obtained. This means that, within our approach, the results
in Eqs.~(\ref{eq:Dxp2})-(\ref{eq:Dpp2}) are only valid for $t \gg
\tau$. In fact, for $t \lesssim \tau$, they essentially coincide with
those of Ref.~\cite{QBM1} for an ohmic environment. For instance, the
term $\mbox{FS}_4(t)$ in Eq.~(\ref{eq:Dpp2}) gives rise to the same
initial jolt, with a width of order $\Lambda^{-1}$ and an amplitude
proportional to $\Lambda$, found in Ref.~\cite{QBM1}. However, as we
discussed above, the results should be valid for times larger than the
switch-on time because the contribution from the switch-on period to
both the master equation coefficients and the evolution of the density
matrix at $t \gg \tau$ is negligible.

It is interesting to note that the coefficients $D_{xp}(t)$ and
$D_{pp}(t)$ both exhibit logarithmic divergences in the limit $\Lambda
\rightarrow \infty$ (for times larger than the switch-on time and thus
larger than $\Lambda^{-1}$) due to the term proportional to
$\mathrm{FI}_3$.\footnote{As it is well-known and can be checked from
  Eq.~(\ref{eq:FI_N}), the term logarithmic in $\Lambda$ is not
  present in $\mathrm{FI}_3$ when considering a fixed finite cut-off
  $\Lambda$ and temperatures much higher than $\Lambda$.} This has
been pointed out for $D_{xp}(t)$ in Ref.~\cite{lombardo}, where the
coefficients of the master equation were calculated perturbatively to
second order in the system-environment coupling constants (linear
order in $\gamma_0$). The fact that there is also a logarithmic
divergence in $D_{pp}(t)$ was not seen in that reference because it
is quartic in the system-environment coupling constants (quadratic in
$\gamma_0$), as it can be seen in Eq.~(\ref{eq:Dpp2}). Moreover, such
kinds of perturbative calculations cannot be employed to study the
long time behavior since they are only valid for $\gamma_0 t \ll 1$
and they miss for instance the exponential decay of the second and
third terms on the right-hand side of
Eqs.~(\ref{eq:Dxp2})-(\ref{eq:Dpp2}).

\subsection{Exact Late-Time Behavior and Approximate High and Low-Temperature
Results}
\label{sec:master-late}

For the late-time diffusion coefficients, only two integrals need to
be performed and they can be expressed in closed form with reasonably
intuitive functions:\footnote{\label{overdamping}Many of the
  expressions derived throughout this paper assume underdamping with
  an extreme cut-off, \emph{i.e.}, $\gamma_0 < \Omega_r < \Lambda$
  with $\tilde{\Omega} = \sqrt{\Omega_r^2 - \gamma_0^2}$. They can be
  used for the overdamping regime by making the following analytical
  continuation: $\tilde{\Omega} \to \imath \tilde{\gamma}$ with
  $\tilde{\gamma} =\sqrt{\gamma_0^2 - \Omega_r^2}$ real.\\
  Therefore, Eqs.~\eqref{eq:FI_1}-\eqref{eq:FI_3} can be applied to
  the overdamping case if the $\mathrm{Im}$ and $\mathrm{Re}$ terms
  are first expanded assuming $\tilde{\Omega}$ is real, and then the
  analytical continuation $\tilde{\Omega} \to \imath \tilde{\gamma}$
  is made.}
\begin{eqnarray}
\mbox{FI}_1 & = & \frac{\pi T}{2 \gamma_0 \Omega_r^2} + \frac{1}{2 \gamma_0 \tilde {\Omega}}
\mbox{Im}\left[ \mbox{H}\left( \frac{\gamma_0 + \imath \tilde{\Omega}}{2 \pi T} \right) \right]
\label{eq:FI_1} , \\
\mbox{FI}_3 & = & \frac{\pi T}{2 \gamma_0} + \frac{\tilde{\Omega}^2-\gamma_0^2}
{2 \gamma_0 \tilde{\Omega}} \mbox{Im}\left[ \mbox{H}\left( \frac{\gamma_0 + \imath \tilde{\Omega}}
{2 \pi T} \right) \right] \notag \\
&& +\mbox{Re}\left[ \mbox{H}\left(\frac{\Lambda}{2 \pi T}\right)
- \mbox{H}\left( \frac{\gamma_0 + \imath \tilde{\Omega}}{2 \pi T} \right)\right] \label{eq:FI_3} ,
\end{eqnarray}
where terms involving negative powers of the cut-off have been
neglected and $\mbox{H}(z)$ is the harmonic number function defined
in Appendix~\ref{sec:harmonic}.  It behaves like $\log{(z)}$ except near
the origin, where it does not diverge and actually vanishes.  The
$\mbox{Re}[\cdots]$ term in $\mbox{FI}_3$ is effectively the
$\log(\Lambda /\Omega_r)$ divergence. It disappears at extreme
temperature (higher than the cut-off).

\subsubsection{Expansion from Zero Temperature}
\label{sec:master-low}

A low-temperature expansion of the master equation coefficients can be
obtained using the following expansion for the hyperbolic cotangent in
the corresponding frequency integrals:
\begin{equation}
\coth\left(\frac{\omega}{2 T}\right) = 1 + 2 \sum_{k=1}^\infty e^{-k \frac{\omega}{T}}
\label{eq:coth_lowT} .
\end{equation}
Moreover, since the results for all the relevant integrals can be
obtained from $\mathrm{FC}_1(t)$ simply by differentiating with
respect to $t$, we will concentrate on computing Eq.~\eqref{eq:FC_N}
with $N=1$. In order to do that, one first decomposes the integrand in
Eq.~\eqref{eq:FC_N} into partial fractions and splits the integral
$\int_0^\Lambda d\omega$ into $\int_0^\infty d\omega -
\int_\Lambda^\infty d\omega$. Finally, using Eq.~\eqref{eq:coth_lowT}
they become a linear combination of exponential integrals. The result
can be written as
\begin{eqnarray}
\mbox{FC}_1(t) & = & \frac{1}{4 \gamma_0 \tilde{\Omega}} \mbox
{Im}\left[ \frac{\mbox{E}_1\left( +(\gamma_0-\imath \tilde{\Omega})t \right)}{e^{-
(\gamma_0-\imath \tilde{\Omega})t}} + \frac{\mbox{E}_1\left( -(\gamma_0-\imath \tilde
{\Omega})t \right)}{e^{+(\gamma_0-\imath \tilde{\Omega})t}} \right] \nonumber \\
&& + \frac{\pi}{4 \gamma_0 \tilde{\Omega}} \cos(\tilde{\Omega} t) e^{-\gamma_0 t} +
\Delta \mbox{FC}_1^\mathrm{PT}(t) + \Delta \mbox{FC}_1^\mathrm{L\Lambda}(t)
\label{eq:low-approx} ,
\end{eqnarray}
where $\mbox{E}_1(z)$ is the exponential integral detailed in
Appendix~\ref{sec:expint}.  The first three terms come from the
$\int_0^\infty d\omega$ integral of the constant term in
Eq.~\eqref{eq:coth_lowT}. On the other hand, $\Delta
\mbox{FC}_1^\mathrm{PT}(t)$, which vanishes at zero temperature, is
the contribution from the $\int_0^\infty d\omega$ integral of the
remaining terms in Eq.~\eqref{eq:coth_lowT}:
\begin{equation}
\Delta \mbox{FC}_1^\mathrm{PT}(t) =  2 \sum_{k=1}^\infty \int_0^\infty \frac{\omega \cos
(\omega t) e^{-k \frac{\omega}{T}}}{\left( \omega^2-\Omega_r^2 \right)^2 + 4
\gamma_0^2 \omega^2} d\omega
\label{eq:pt} .
\end{equation}
Finally, the term $\Delta \mbox{FC}_1^\mathrm{L\Lambda}(t)$ contains
all the dependence on the cut-off $\Lambda$, which arises from all the
$\int_\Lambda^\infty d\omega$ integrals. In particular, its value for
zero temperature, which corresponds to taking only the first term in
Eq.~\eqref{eq:coth_lowT}, can be calculated explicitly:
\begin{eqnarray}
\Delta \mbox{FC}_1^{Z\Lambda}(t) = -\frac{1}{8 \gamma_0 \tilde{\Omega}} \mbox
{Im} \!\!\!\!\! && \left[ \frac{\mbox{E}_1\left( +(\gamma_0-\imath (\tilde{\Omega}+\Lambda))t
\right)}{e^{-(\gamma_0-\imath \tilde{\Omega})t}} + \frac{\mbox{E}_1\left( -(\gamma_0-\imath
(\tilde{\Omega}+\Lambda))t \right)}{e^{+(\gamma_0-\imath \tilde{\Omega})t}} \right.  \notag \\
&&\,\,\, \left. + \frac{\mbox{E}_1\left( +
(\gamma_0-\imath (\tilde{\Omega}-\Lambda))t \right)}{e^{-(\gamma_0-\imath \tilde
{\Omega})t}} + \frac{\mbox{E}_1\left( -(\gamma_0-\imath (\tilde{\Omega}-\Lambda))t
\right)}{e^{+(\gamma_0-\imath \tilde{\Omega})t}} \right] .
\end{eqnarray}
Note that, since $\mathrm{FC}_1(t)$ is finite in the limit $\Lambda
\to \infty$, the contribution from $\Delta
\mbox{FC}_1^\mathrm{L\Lambda}(t)$ will be very small for sufficiently
large $\Lambda$. However, when differentiating several times with
respect to $t$, this term gives rise to the divergences that
$\mathrm{FC}_3(t)$ and $\mathrm{FS}_4(t)$ exhibit during a short
initial period of order $\Lambda^{-1}$, which are responsible for the
initial jolt of the master equation coefficients found in
Ref.~\cite{QBM1}. Moreover, even in those cases it would be enough to
consider just $\Delta \mbox{FC}_1^\mathrm{Z\Lambda}(t)$ rather than
$\Delta \mbox{FC}_1^\mathrm{L\Lambda}(t)$ because the integral in
Eq.~\eqref{eq:pt} is convergent for arbitrary positive powers of the
frequency due to the exponential factor.

In fact, given that the divergent behavior for short times is
irrelevant when considering a switch-on time sufficiently longer than
$\Lambda^{-1}$, it turns out that in our case we can completely
neglect the contribution from the cut-off dependent terms for
sufficiently large $\Lambda$. It would only be necessary when
obtaining $\mathrm{FI}_3$ by evaluating $\mathrm{FC}_3(t)$ at $t=0$,
but it is much simpler and more accurate in general to use the exact
result for $\mathrm{FI}_1$ and $\mathrm{FI}_3$ computed earlier and
given by Eqs.~\eqref{eq:FI_1}-\eqref{eq:FI_3}.

To sum up, taking into account the remarks in the previous paragraph
together with the fact that the contribution from the third term in
Eq.~\eqref{eq:low-approx} cancels out when adding all the
time-dependent terms in Eqs.~\eqref{eq:Dxp2}-\eqref{eq:Dpp2}, all that
one needs [in addition to Eqs.~\eqref{eq:FI_1}-\eqref{eq:FI_1}] to
calculate the master equation coefficients is
\begin{eqnarray}
\Delta \mbox{FC}_1(t) &=& \frac{1}{4 \gamma_0 \tilde{\Omega}}
\mbox{Im}\left[ \frac{\mbox{E}_1\left( +(\gamma_0-\imath \tilde{\Omega})t \right)}
{e^{-(\gamma_0-\imath \tilde{\Omega})t}} + \frac{\mbox{E}_1\left( -(\gamma_0-\imath \tilde{\Omega})t
\right)}{e^{+(\gamma_0-\imath \tilde{\Omega})t}} \right] \notag \\
&& +2 \sum_{k=1}^\infty \int_0^\infty \frac{\omega \cos(\omega t) e^{-k \frac{\omega}{T}}}
{\left( \omega^2-\Omega_r^2 \right)^2 + 4 \gamma_0^2 \omega^2} d\omega
\label{eq:low-approx2} .
\end{eqnarray}
Given the asymptotic behavior of $\mbox{E}_1(z)$ for large $|z|$, one
can see that the $\mbox{Im}[\cdots]$ terms decay only weakly in time like $1 /
(\Omega_\mathrm{r} t)^2$; they both have complicated $\tilde{\Omega}$ frequency
oscillations. The integral in the last term, which vanishes when the
temperature tends to zero, can be performed explicitly, but the sum
becomes particularly complicated.

\subsubsection{Asymptotic Expansion from High Temperature}
\label{sec:master-high}

A high-temperature expansion can be obtained by making use of the
following expression for the hyperbolic cotangent:
\begin{equation}
\coth\left(\frac{\omega}{2 T}\right) = \frac{2 T}{\omega} + \sum_{k=1}^\infty \frac{4 T
\omega}{\omega^2 + (2 \pi T k)^2} ,
\label{eq:highTcoth}
\end{equation}
which will yield a series solution that is best in the high
temperature, late time regime $2 \pi T t \gg 0$. $\mbox{FC}_1(t)$ can
then be calculated in several steps.

First, one calculates the $\int_0^\infty d\omega$ integral
corresponding to Eq.~\eqref{eq:FC_N} considering only the first term
in Eq.~\eqref{eq:highTcoth}, decomposes the resulting integrand into
partial fractions and ends up with a linear combination of Exponential
Integrals. After simplification, the result can be written as
\begin{equation}
\mbox{FC}_1^\mathrm{HT}(t) = \frac{\pi  T}{2 \tilde{\Omega} \Omega _r^2  \gamma _0}
\left(\tilde{\Omega} \cos (\tilde{\Omega}t) + \gamma _0 \sin (\tilde{\Omega}t\right)
e^{-\gamma _0 t} .
\end{equation}
Next, we perform the $\int_0^\infty d\omega$ integral corresponding to
Eq.~\eqref{eq:FC_N} with the general term in the sum of
Eq.~\eqref{eq:highTcoth} replacing the hyperbolic cotangent. After
decomposing into partial fractions we end up again with a linear
combination of Exponential Integrals, and the final result after
simplification is
\begin{eqnarray}
\mbox{FC}_1^\mathrm{LT}(t) &=& \sum_{k=1}^\infty \,
\frac{ \pi  T e^{-\gamma _0 t}} {\tilde{\Omega} \gamma _0
\left(\tilde{\Omega}^4+2 \left(4 k^2 \pi ^2 T^2+\gamma _0^2\right)
\tilde{\Omega}^2+\left(\gamma _0^2-4 k^2 \pi ^2 T^2\right)^2\right)}
\left[ -4 k \pi  T\tilde{\Omega}\gamma _0 e^{(\gamma _0 - 2 k\pi  T) t}
 \right. \nonumber \\
&&\left. +  \gamma _0 \left(-4 k^2 \pi ^2 T^2+\tilde{\Omega }^2+\gamma _0^2\right)
\sin(\tilde{\Omega}t) +  \tilde{\Omega} \left(4 k^2 \pi^2 T^2+\tilde{\Omega }^2
+\gamma _0^2\right) \cos(\tilde{\Omega}t)\right] .
\end{eqnarray}
The sum over $k$ can be performed explicitly for the terms multiplying
the sine and the cosine, so that we get
\begin{equation}
\mbox{FC}_1^\mathrm{LT}(t) =  \frac{\pi}{4 \gamma_0 \tilde{\Omega}}
\frac{\sinh\left(\frac{\tilde{\Omega}}{T}\right) \cos(\tilde{\Omega} t)
+ \sin\left(\frac{\gamma_0}{T}\right) \sin(\tilde{\Omega} t)}
{\cosh\left(\frac{\tilde{\Omega}}{T}\right)-\cos\left(\frac{\gamma_0}
{T}\right)} e^{-\gamma_0 t} + \Delta \mbox{FE}_1^\mathrm{LT}(t)
\label{eq:htfc1} ,
\end{equation}
with
\begin{equation}
\Delta \mbox{FE}_N^\mathrm{LT}(t) = \frac{1}{(-2 \pi T)^2} \sum_{k=1}^{\infty} \frac{k e^{-2\pi T t k}}
{(k^2+\left(\frac{\Omega_r}{2\pi T}\right)^2)-4\left(\frac{\gamma_0}{2\pi T}\right)^2 k^2} ,
\end{equation}
which satisfies the following property:
\begin{equation}
\frac{d}{dt} \Delta \mbox{FE}_N^\mathrm{LT}(t) = -(2\pi T) \Delta \mbox{FE}_{N+1}^\mathrm{LT}(t) .
\end{equation}

Putting everything together, we have
\begin{eqnarray}
\mbox{FC}_1(t) & = &  \frac{\pi}{4 \gamma_0 \tilde{\Omega}} \frac{\sinh\left
(\frac{\tilde{\Omega}}{T}\right) \cos(\tilde{\Omega} t) + \sin\left(\frac{\gamma_0}{T}\right)
\sin(\tilde{\Omega} t)}{\cosh\left(\frac{\tilde{\Omega}}{T}\right)-\cos\left(\frac{\gamma_0}
{T}\right)} e^{-\gamma_0 t} \notag \\
&& + \Delta \mbox{FE}_1^\mathrm{LT}(t) + \Delta \mbox{FC}_1^ \mathrm{H\Lambda}(t)
\label{eq:htfc2} ,
\end{eqnarray}
where $\Delta \mbox{FC}_1^ \mathrm{H\Lambda}(t)$ contains all the
dependence on the cut-off $\Lambda$, and comes from all the
$\int_\Lambda^\infty d\omega$ integrals. Nevertheless, for the same
reasons given in the previous subsection, in our case one can neglect
this term for sufficiently large values of $\Lambda$. Furthermore,
since the contribution from the first term in Eq.~\eqref{eq:htfc2}
cancels out when adding all the time-dependent terms in
Eqs.~\eqref{eq:Dxp2}-\eqref{eq:Dpp2}, and provided that we use
Eqs.~\eqref{eq:FI_1}-\eqref{eq:FI_3} to compute $\mathrm{FI}_1$ and
$\mathrm{FI}_3$, it is sufficient to consider
\begin{equation}
\Delta \mbox{FC}_1(t) = \frac{1}{(-2 \pi T)^2} \sum_{k=1}^{\infty} \frac{k e^{-2\pi T t k}}
{(k^2+\left(\frac{\Omega_r}{2\pi T}\right)^2)-4\left(\frac{\gamma_0}{2\pi T}\right)^2 k^2}
\label{eq:high-approx1} ,
\end{equation}
in order to calculate the master equation coefficients.

All terms on the right-hand side of Eq.~\eqref{eq:high-approx1} decay
with temperature-dependent time-scales. Since it cannot be expressed
in closed form with any intuitive functions, the sum is not
explicitly performed. Although useful for numerical evaluation, this
form does not immediately reveal the $\gamma_0 t$ and $\tilde{\Omega}
t$ behavior. Therefore, it is convenient to approximate the sum by an
integral that can be computed explicitly, as follows:
\begin{eqnarray}
\sum_{k=1}^\infty & \approx & \int_1^\infty dk \notag , \\
\label{eq:high-approx} \Delta\mbox{FC}_1(t) & \approx & \frac{1}{4 \gamma_0 \tilde{\Omega}}
\mbox{Im}\left[  \frac{ \mbox{Ei}\left((\gamma_0-\imath \tilde{\Omega} - 2\pi T)t \right)}
{e^{(\gamma_0-\imath \tilde{\Omega})t}} + \frac{ \mbox{Ei}\left((-\gamma_0+\imath \tilde{\Omega}
- 2\pi T)t \right)}{e^{(-\gamma_0+\imath \tilde{\Omega})t}} \right]
\label{eq:high-approx2}.
\end{eqnarray}
{From} this expression one can see that the most lingering terms decay
exponentially as $e^{-2\pi T t} \, \mathcal{O}(1/t)$.
There are also oscillations with frequency $\tilde{\Omega}$.

\subsubsection{Approximate General Solution}
\label{sec:master-approx}

By inspecting the corresponding terms in the high
\eqref{eq:high-approx2} and low-temperature \eqref{eq:low-approx2}
expansions, one can get a good idea of the general solution:
\begin{eqnarray}
\mbox{FC}_1(t) & \approx & \frac{\pi}{4 \gamma_0 \tilde{\Omega}} \frac{\sinh\left(\frac
{\tilde{\Omega}}{T}\right) \cos(\tilde{\Omega} t) + \sin\left(\frac{\gamma_0}{T}\right) \sin
(\tilde{\Omega} t)}{\cosh\left(\frac{\tilde{\Omega}}{T}\right)-\cos\left(\frac{\gamma_0}{T}
\right)} e^{-\gamma_0 t} \notag \\
&& + \frac{1}{4 \gamma_0 \tilde{\Omega}} \mbox{Im}\left[  \frac{ \mbox{E}_1\left((-
\gamma_0+\imath \tilde{\Omega} + 2\pi T)t \right)}{e^{(\gamma_0-\imath \tilde{\Omega})
t}} + \frac{ \mbox{E}_1\left((\gamma_0-\imath \tilde{\Omega} + 2\pi T)t \right)}{e^{(-
\gamma_0+\imath \tilde{\Omega})t}} \right] \notag\\
&& + \Delta \mbox{FC}_1^{\mathrm{H}\Lambda} (t) .
\end{eqnarray}
This expression is exact at both zero temperature and high
temperature, and it has only minor discrepancies within a small range
of intermediate temperatures.  Moreover, as explained above, all that
one really needs to calculate the master equation coefficients is
\begin{equation}
\Delta \mbox{FC}_1(t) \approx \frac{1}{4 \gamma_0 \tilde{\Omega}}
\mbox{Im}\left[  \frac{ \mbox{E}_1\left((-\gamma_0+\imath \tilde{\Omega} + 2\pi T)t \right)}
{e^{(\gamma_0-\imath \tilde{\Omega})t}} + \frac{ \mbox{E}_1\left((\gamma_0-\imath \tilde{\Omega}
+ 2\pi T)t \right)}{e^{(-\gamma_0+\imath \tilde{\Omega})t}} \right] .
\end{equation}

\subsubsection{Comparison with Numerical Results}

{From} our analysis of the integrals it can be seen that there is only
one fluctuation frequency $\tilde{\Omega}$ and it is always damped.
There are only two damping rates, $\gamma_0$ and $2 \pi T$.  All
temperature-damped terms also have coupling damping, so one can say
that $t \gg \gamma_0^{-1}$ is always late time, although the
temperature can hasten this.

Since we have solved the integrals in the different temperature
regimes, we will now plot the lowest order terms of each expansion in
the different temperature regimes.  The coupling will be set at a
moderate level to enhance their differences.

\begin{figure}[h]
\centering
\includegraphics[width=4in]{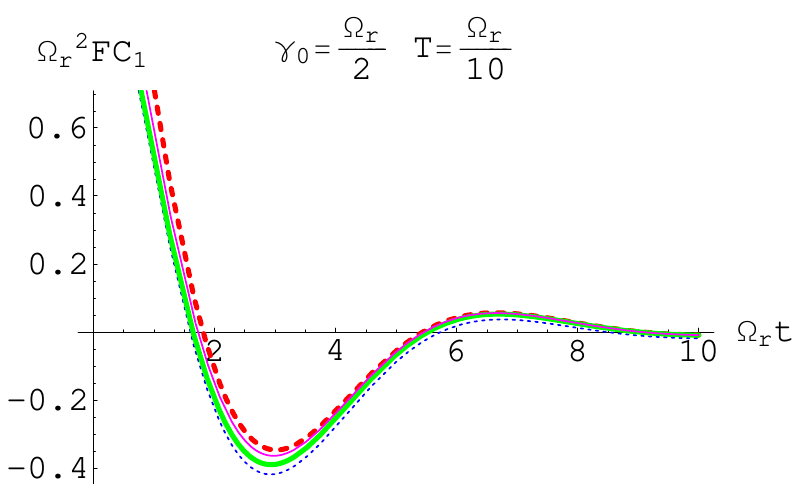}
\caption{The case of a moderate temperature.
$\mbox{FC}_1$ calculated \textcolor{green}{\boldmath$-$ \textbf{numerically}}, to first
order in the \textcolor{blue}{$\cdots$ low-temperature expansion}, to first order in the
\textcolor{red}{\boldmath$\cdots$ \textbf{high-temperature expansion}}, and with the
\textcolor{magenta}{$-$ approximate general solution}.
All of the approximate solutions are fairly close with some minor discrepancy at short
times. The low-temperature expansion will continue to be slightly off into late times.}
\end{figure}

\begin{figure}[h]
\centering
\includegraphics[width=4in]{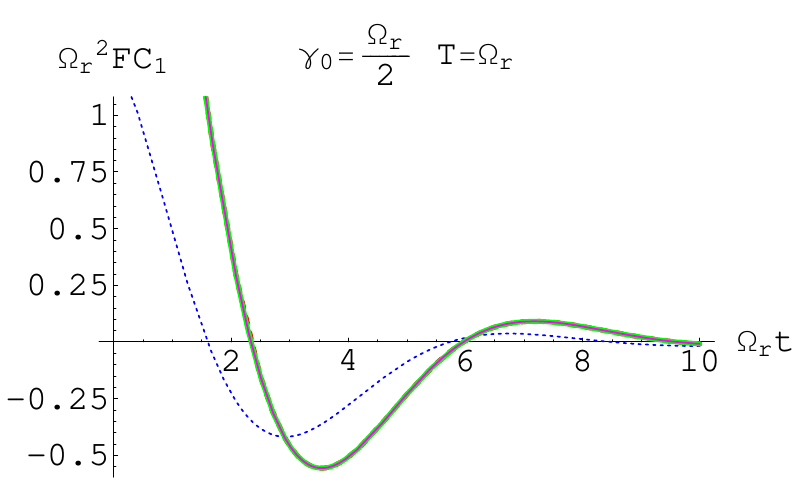}
\caption{The case of a high temperature.
The \textcolor{magenta}{$-$ approximate general solution} and the \textcolor{red}
{\boldmath$\cdots$ \textbf{first order high-temperature solution}} stick very well to the
\textcolor{green}{\boldmath$-$ \textbf{numeric solution}}, while the \textcolor{blue}{$
\cdots$ first order low-temperature solution} has amplitude and phase discrepancies.}
\end{figure}

Note that the high-temperature expansion (including the first order
correction) gives very good agreement even at much lower temperatures
$T \sim \Omega_r$ (much lower than the regime $T > \Lambda$ where the
high-temperature expansion is clearly expected to be very accurate).

\begin{figure}[h]
\centering
\includegraphics[width=4in]{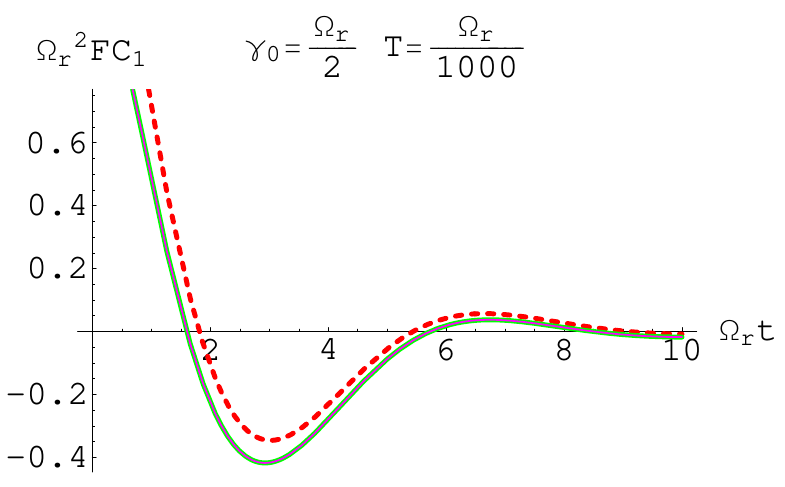}
\caption{The case of a low temperature.
The \textcolor{magenta}{$-$ approximate general solution} and the \textcolor{blue}{$
\cdots$ first order low temperature solution} stick very well to the \textcolor{green}
{\boldmath$-$ \textbf{numeric solution}}, while the \textcolor{red}{\boldmath$\cdots$
\textbf{first order high temperature solution}} has some amplitude and phase discrepancy
initially.}
\end{figure}

When we say $n^\mathrm{th}$ order in the high-temperature expansion,
we more acurately mean up to the $e^{-2 \pi T t n}$ term in the
high-temperature expansion.  Strictly speaking this is not a
temperature expansion but an expansion in $e^{-2 \pi T t}$, the only
terms which decay at a temperature-dependent rate, so it is really a
high-temperature, late-time expansion.  Similarly, when we say
$n^\mathrm{th}$ order in the low-temperature expansion, we mean
keeping the $n^\mathrm{th}$ order term in the sum appearing in the
last term of Eq.~\eqref{eq:low-approx2}.

\subsection{Some Closed Form Results}

\subsubsection{Extreme Temperature, Arbitrary Time}

Considering just the first term on the right-hand side of
Eq.~\eqref{eq:highTcoth} for the hyperbolic cotangent, one gets the
following result for temperatures much higher than the cut-off:
\begin{eqnarray}
T & \gg & \Lambda , \notag\\
D_{xp} & = & 0 , \\
D_{pp} & = & +2 M \gamma_0 T .
\end{eqnarray}
This is the common, but questionable high-temperature limit.  There is
no first order term for the anomalous diffusion coefficient, but there
are lower order terms that vanish only for infinite temperature.  They
include the term that gives rise to the $\log{\Lambda}$ divergence
when the temperature becomes much smaller than $\Lambda$.

\subsubsection{Arbitrary Temperature, Late Time}
\label{sec:latediff}

Neglecting the terms multiplied by the factor $e^{-\gamma_0 t}$ in
Eqs.~\eqref{eq:Dxp2}-\eqref{eq:Dpp2}, we obtain the following result,
which is valid for times much larger than the relaxation time
(\emph{i.e.}, $t \gg \gamma_0^{-1}$):
\begin{eqnarray}
t & \gg & \frac{1}{\gamma_0} , \notag\\
D_{xp} & = & -\frac{\gamma_0^2}{\pi \tilde{\Omega}} \mbox{Im}\left[ \mbox{H} \left( \frac
{\gamma_0 + \imath \tilde{\Omega}}{2 \pi T} \right) \right] \notag \\
&& + \frac{\gamma_0}{\pi} \mbox{Re}\left[ \mbox{H}\left(\frac{\Lambda}{2 \pi T}\right) -
\mbox{H}\left( \frac{\gamma_0 + \imath \tilde{\Omega}}{2 \pi T} \right)\right]
\label{eq:lateDxp} ,\\
D_{pp} & = & +2 M \gamma_0 T + \frac{2 M \gamma_0}{\pi \tilde{\Omega}} ( \tilde
{\Omega}^2 - \gamma_0^2 ) \mbox{Im}\left[ \mbox{H} \left( \frac{\gamma_0 + \imath \tilde
{\Omega}}{2 \pi T} \right)\right] \notag \\
&& + \frac{4 M \gamma_0^2}{\pi} \mbox{Re}\left[ \mbox{H}\left(\frac{\Lambda}{2 \pi T}
\right) - \mbox{H}\left( \frac{\gamma_0 + \imath \tilde{\Omega}}{2 \pi T} \right)\right]
\label{eq:lateDpp} .
\end{eqnarray}
The $\mathrm{Re}[\cdots]$ terms become $\log(\Lambda / \Omega_r)$ at
zero temperature and vanish in the infinite temperature limit. On the
other hand, the value of the $\mathrm{Im}[\cdots]$ terms is
$\cos^{-1}(\gamma_0 / \Omega_r)$ at zero temperature, whereas it also
vanishes in the infinite-temperature limit.

\subsection{Comparison with Caldeira \emph{et al.}}
\label{sec-caldeira}

We now compare our results to CCR's weak coupling master equation
\cite{caldeira} , which differs from the HPZ equation and has the
following diffusion coefficients:
\begin{eqnarray}
D_{xp}^{C} & = & 0 , \\
D_{pp}^{C} & = & \gamma_0 M \Omega_r \coth{\left( \frac{\Omega_r}{2T} \right)} .
\end{eqnarray}
The CCR master equation is frequently used for its simplicity and is
believed to be accurate at late times for weak coupling and arbitrary
temperature.
\begin{figure}[h]
\centering
\includegraphics[width=4in]{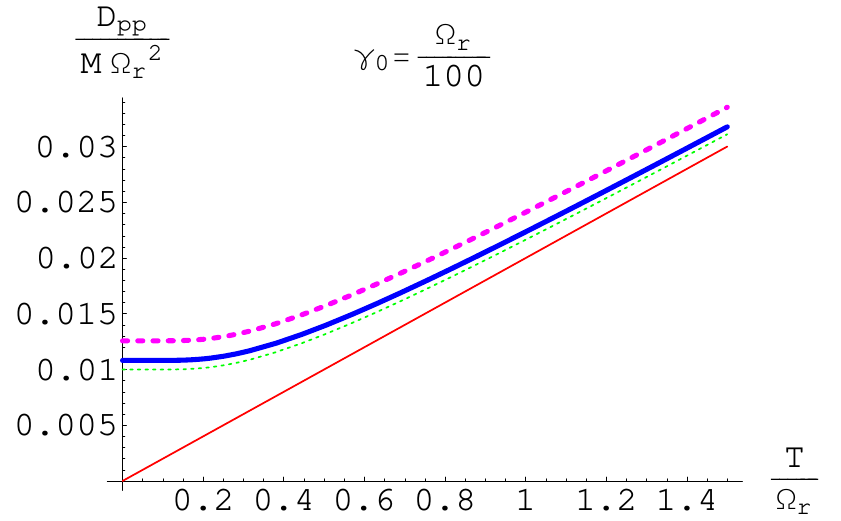}
\caption{Late time $D_{pp}$ for \textcolor{red}{$-$ high temperature}, \textcolor{green}{$
\cdots$ Caldeira}, \textcolor{blue}{\boldmath$-$ \textbf{HPZ at} $\Lambda=10^3
\Omega_r$}, and \textcolor{magenta}{\boldmath$\cdots$ \textbf{HPZ at} $\Lambda=10^9
\Omega_r$}.}
\label{fig:CCR1}
\end{figure}
One can see from Fig.~\ref{fig:CCR1} that, ignoring the contribution
from the cut-off, the CCR approximation matches extremely well with
our exact results for the \emph{normal diffusion coefficient} at weak
coupling.  The CCR error from neglecting the cut-off dependence is
determined by the order of magnitude of the cut-off scale since it is
logarithmic in $\Lambda$.  In any case, the CCR approximation
underestimates the magnitude of the diffusion coefficient.
\begin{figure}[h]
\centering
\includegraphics[width=4in]{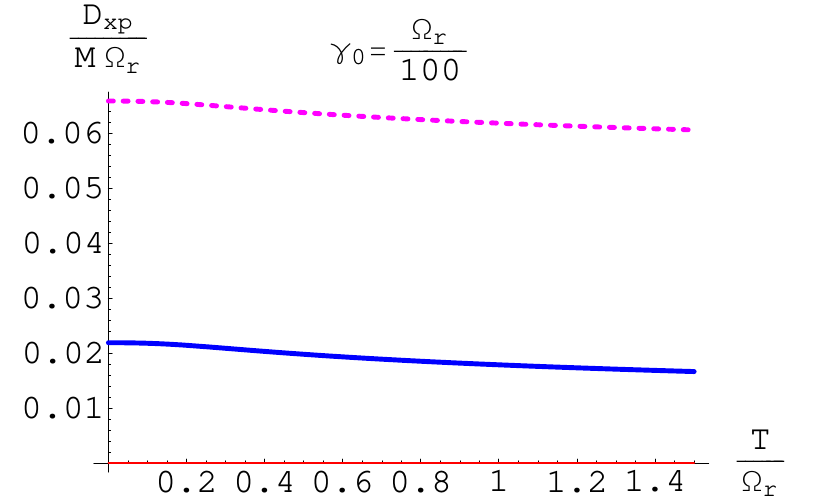}
\caption{Late time $D_{xp}$ for \textcolor{red}{$-$ high temperature or equivalently
Caldeira}, \textcolor{blue}{\boldmath$-$ \textbf{HPZ at} $\Lambda=10^3 \Omega_r$},
and \textcolor{magenta}{\boldmath$\cdots$ \textbf{HPZ at} $\Lambda=10^9 \Omega_r
$}.}
\label{fig:CCR2}
\end{figure}

Unlike the normal diffusion coefficient, the difference is much more
severe for the \emph{anomalous diffusion coefficient}, which is
\emph{completely absent} in CCR.  The largest contribution (in the
weak coupling regime) to the anomalous diffusion coefficient comes
from the cut-off and it does not vanish at finite temperature (see
Fig.~\ref{fig:CCR2}). Moreover, it cannot be regarded as vanishing at
weak coupling because it is only proportional to one power of the
coupling constant, which is the order to which CCR's master equation
should be valid.

The final point that should be made is that CCR's arbitrary
temperature master equation is only valid in the weak coupling regime
for late times and, unfortunately, late time always means $t \gg
\gamma_0^{-1}$ (according to our exact results for the HPZ master
equation). Therefore, the weaker the coupling, the longer one must
wait for CCR's master equation to be valid.

\section{Master Equation Coefficients with Laurent Series Spectra }
\label{sec:master-laurent}

\subsection{Derivation of The General Integrals}
\label{sec:master_general}

We start by considering the independent effect of power-law terms in
the spectral function with ultraviolet cut-off $\Lambda$ and infrared
cut-off $\lambda$ whenever necessary. We take these cut-offs to be
extremely large and extremely small respectively, not because that
should necessarily be the case for any relevant physical situation,
but to restrict the model to something that we can solve
explicitly. Hence, we take a spectral function of the form
\begin{eqnarray}
I_N(\omega) = \frac{2}{\pi} M \gamma_{N-1} \omega \left( \frac{\omega}{\omega_c}
\right)^{N-1} && (\lambda < \omega < \Lambda) ,
\end{eqnarray}
where $N$ is an integer number and $\omega_c^{|N-1|}$ is a product of
characteristic frequencies chosen so that well-behaved results are
obtained for the master equation coefficients in the limits $\Lambda
\to \infty$ and $\lambda \to 0$. Traditionally $\omega_c$ has been
chosen to be $\Lambda$ whenever a UV cut-off was necessary; we will
end up making a similar choice here.

\subsubsection{The Classical Trajectories}

Given a general spectral function the equation of motion for the
system trajectories corresponds to an integro-differential equation:
\begin{equation}
0 = \ddot{u}(s) + \Omega^2 u(s) + \frac{2}{M} \int_0^s \mu(s-s') u(s') ds' .
\end{equation}
It is thus convenient to perform a Laplace transform
\begin{equation}
\hat{f}(\zeta) = \int_0^\infty e^{-s \zeta} f(s) ds ,
\end{equation}
under which the equation becomes purely algebraic, since the integral
involving the dissipation kernel is merely a convolution. We then have
\begin{equation}
\zeta u_0 + \dot{u}_0 = \left( \zeta^2 + \frac{2}{M} \hat{\mu}(\zeta) + \Omega^2 \right)
\hat{u}(\zeta) ,
\end{equation}
or, equivalently,
\begin{equation}
\hat{u}(\zeta) = \frac{\zeta u_0 + \dot{u}_0}{\zeta^2 + \frac{2}{M} \hat{\mu}(\zeta) +
\Omega^2 } \, ,
\end{equation}
where $u_0, \dot{u}_0$ are the initial conditions for $u(s)$.
Given the expression of the dissipation kernel
\begin{equation}
\mu(s) = - \int_\lambda^\Lambda I_N(\omega) \sin{(\omega s)} d\omega ,
\end{equation}
one can easily compute its Laplace transform:
\begin{equation}
\hat{\mu}(\zeta) = \frac{2}{\pi} M \gamma_{N-1} \int_\lambda^\Lambda \frac
{\omega^2 }{\omega^2 + \zeta^2} \left( \frac{\omega}{\omega_c} \right)^{N-1} d\omega
\label{eq:lapl_dissip}.
\end{equation}
The results for different integer values of $N$ are listed in
Table~\ref{tab:intdiss}. The column labeled with $\hat{\mu}_c(\zeta)$
corresponds to the outcome of the integral in
Eq.~\eqref{eq:lapl_dissip} after taking the limits $\Lambda \to
\infty$ and $\lambda \to 0$ wherever that gave a finite result. The
values of $\omega_c$ were chosen so that the integrals that appear
when calculating the diffusion coefficients were finite (more details
can be found in the next two subsections). Finally, the column labeled
with $\hat{\mu}_c(\zeta)$ follows from substituting $\omega_c$ in
$\hat{\mu}_c(\zeta)$ with the values of the previous column and taking
the limits $\Lambda \to \infty$ and $\lambda \to 0$ whenever they give
a finite result.

Had we chosen fractional powers of the frequency in the spectrum or
even a different functional dependence such as a logarithm, this
would have resulted in an integro-differential equation for the
classical system trajectories.  In addition to being less classical,
they would also be much more difficult to solve.

\begin{table}[h]
\centering
\begin{tabular}{|c|c|c|c|}
\hline
$N-1$ & $\hat{\mu}_c(\zeta)$ & $\omega_c^{|N-1|}$ & $\hat{\mu}(\zeta)$\\
\hline
$-n$&$\cdots$   &   $\lambda^{n-1} \omega_c$    &   $0$\\
$-4$&   $M \gamma_{-4} \left( -\frac{2}{\pi}\frac{\omega_c^4}{\zeta^2} \frac{1}
{\lambda} + \frac{\omega_c^4}{\zeta^3} \right)$ &   $\lambda^3 \omega_c$    &
$0$\\
$-3$&   $\frac{2}{\pi} M \gamma_{-3} \left( -\frac{\omega_c^3}{2 \zeta^2} \log{\frac
{\lambda}{\zeta}} \right) $ & $\lambda^2 \omega_c$  &   $0$\\
$-2$&   $M \gamma_{-2} \left( - \frac{\omega_c^2}{\zeta} \right)$   &
$\lambda\omega_c$   &  $0$\\
$-1$    &   $\frac{2}{\pi} M \gamma_{-1} \left( -\omega_c \log{\frac{\Lambda}{\zeta}} \right)
$ & $\omega_c$  &   $\frac{2}{\pi} M \gamma_{-1} \left( -\omega_c \log
{\frac{\Lambda}{\zeta}} \right)$\\
$0$ &   $M \gamma_0 \left( \zeta - \frac{2}{\pi} \Lambda \right)$
&   $1$ &   $M
\gamma_0 \left( \zeta - \frac{2}{\pi} \Lambda \right)$\\
$1$ &   $\frac{2}{\pi} M \gamma_1 \left( \frac{\zeta^2}{\omega_c} \log{\frac{\Lambda}
{\zeta}} - \frac{\Lambda^2}{2 \omega_c} \right)$ & $\Lambda$    &   $-\frac{2}{\pi} M
\gamma_1 \frac{\Lambda}{2}$\\
$2$ &   $M \gamma_2 \left( -\frac{\zeta^3}{\omega_c^2} + \frac{2}{\pi} \frac{\Lambda}
{\omega_c^2} \zeta^2 - \frac{2}{\pi} \frac{\Lambda^3}{3 \omega_c^2} \right)$    &   $
\Lambda^2$  &   $-\frac{2}{\pi} M \gamma_2 \frac{\Lambda}{3}$\\
$3$ &   $\frac{2}{\pi} M \gamma_3 \left( -\frac{\zeta^4}{\omega_c^3} \log{\frac
{\Lambda}{\zeta}} + \frac{\Lambda^2}{2 \omega_c^3} \zeta^2 - \frac{\Lambda^4}{4
\omega_c^3} \right)$ &  $\Lambda^3$ &   $-\frac{2}{\pi} M \gamma_3 \frac{\Lambda}
{4}$\\
$n$ &   $\cdots$    &   $\Lambda^{n}$   &   $-\frac{2}{\pi} M \gamma_{n}
\frac{\Lambda}{n+1}$\\
\hline
\end{tabular}
\caption{\label{tab:intdiss}Laplace transform of the dissipation kernels for various
power-law spectral functions after discarding $\mathcal{O}(\lambda)$ and
$\mathcal{O}(1/\Lambda)$ terms.
$N$ is the power of the spectrum: $N=1$ is ohmic, $N>1$ is supraohmic and $N<1$ is subohmic.
$\hat{\mu}_c(\zeta)$ is the Laplace transform of the dissipation kernel before a
reasonable choice of characteristic frequencies $\omega_c^{|N-1|}$ is made, whereas
$\hat{\mu}(\zeta)$ is the choosen physical dissipation kernel.
Constant terms are renormalizations to the bare frequency.}
\end{table}

Integer \emph{subohmic} spectra, (corresponding to $N-1<0$ in
Table~\ref{tab:intdiss}) all exhibit a similar behavior, except for
the $N=0$ case. For $N<0$ it is never possible to change the classical
ordinary differential equation (ODE) while keeping the diffusion
coefficients finite in the limit $\lambda \to 0$. Indeed, the only way
to guarantee that all the frequency integrals involving the noise
kernel that appear when calculating the diffusion coefficients are
finite is by taking $\omega_c^{n} \sim \lambda^{n-1}$ (higher powers
of $\lambda$ are allowed, but would give vanishing results for the
diffusion coefficients). In that case, there are no additional
contributions to the equation of motion left. In contrast, for $N=0$
the diffusion coefficients only exhibit a logarithmic IR divergence
(proportional to $\log \lambda$) when considering a finite
non-vanishing value of $\omega_c$ independent of $\lambda$. Since we
tolerated logarithmic dependences on the UV cut-off $\Lambda$ in the
ohmic case, it would be natural to allow a similar situation with the
IR cut-off $\lambda$. However, in that case the Laplace transform of
the dissipation kernel exhibits a logarithmic term that makes the
equation of motion for the system trajectories (which corresponds to
an integro-differential equation) much more difficult to
solve. Therefore, we will not consider the $N=0$ case, for the same
reason why we did not consider fractional powers of the frequency in
the spectral function or even a non-logarithmic dependence on the
frequency.

On the other hand, integer \emph{supraohmic} spectra (corresponding to
$N-1>0$ in Table~\ref{tab:intdiss}) all exhibit a similar
behavior. None of them are capable of changing the nature of the
classical ODE while keeping the diffusion coefficients finite in the
limit $\Lambda \to \infty$. Indeed, the only way to guarantee that all
the frequency integrals involving the noise kernel that appear when
calculating the diffusion coefficients are finite is by taking
$\omega_c^{n} \sim \Lambda^{n}$ (higher powers of $\Lambda$ are
allowed, but would give vanishing results for the diffusion
coefficients). In that case, the only contributions to the equation of
motion that are left are frequency renormalization terms.

\subsection{Analytic Spectra Master Equation Coefficients}
\label{sec:master-power}

In this subsection we consider spectral functions that are analytic in
the frequency (and vanish in the limit of zero frequency). They have
the following form:
\begin{equation}
I_{\{\gamma\}}(\omega) = \frac{2}{\pi} M \omega \gamma\left( \frac{\omega}{\Lambda}
\right) \mbox{~~} (\omega < \Lambda)
\label{eq:analytic_spec1},
\end{equation}
where $\gamma$ is an analytic function. Eq.~\eqref{eq:analytic_spec1}
can be written as
\begin{equation}
I_{\{\gamma\}}(\omega) = \frac{2}{\pi} M \omega \sum_{n=0}^{\infty} \gamma_n \left(
\frac{\omega}{\Lambda} \right)^n \mbox{~~} (\omega < \Lambda)
\label{eq:analytic_spec2},
\end{equation}
which corresponds to a linear combination of terms with $N \geq 1$
among those considered in the previous subsection.



\subsubsection{The Classical Trajectories}

The classical equation of motion associated with these spectra is an
ODE that differs trivially from the ohmic case:
\begin{equation}
\zeta u_0 + \dot{u}_0 = \zeta^2 \hat{u}(\zeta) + \frac{2}{M} \left( M \gamma_0 \zeta -
\frac{2}{\pi} M \sum_{n=0}^{\infty} \frac{\gamma_n}{n+1} \Lambda \right) \hat{u}(\zeta) +
\Omega^2 \hat{u}(\zeta) ,
\end{equation}
which can be written in the more compact form
\begin{equation}
\zeta u_0 + \dot{u}_0 = \left( \zeta^2 + 2 \gamma_0 \zeta + \Omega_r^2 \right) \hat{u}
(\zeta) ,
\end{equation}
by introducing the renormalized frequency
\begin{equation}
\Omega_r^2 = \Omega^2 - \frac{4}{\pi} \Lambda \sum_{n=0}^{\infty} \frac{\gamma_n}
{n+1} .
\end{equation}
We can see that only the $\omega^1$ power (from the ohmic-like term)
gives rise to the dissipation term,\footnote{Note that, since an
  additional term $u_0$ is missing, the inverse Laplace transform of
  $2 \gamma_0 \zeta \hat{u}(\zeta)$ gives not only the local
  dissipation term $2 \gamma_0 \dot{u}(t)$, but also a term
  proportional to $\delta(t) u(0)$, which corresponds to the term on
  the right-hand side of Eq.~\eqref{eq:eom1} and is responsible for
  the initial kick when no smooth switching-on function is present.}
whereas all the remaining supraohmic terms merely contribute to an
additional renormalization of the bare frequency.  In fact, one could
rig supraohmic anticoupling terms to eradicate the divergent
difference between the bare and renormalized frequency, although there
is no physical motivation to do so.

\subsubsection{Master Equation Coefficients}

For times longer than the switch-on time, when our explicit
expressions for the diffusion coefficients can be used, the only
difference between the master equation for an analytic spectrum and
the ohmic spectrum lies in a modification of the contribution from the
$\mathrm{FI}_3$ integrals to the diffusion coefficients. This can be
seen as follows. When considering a term proportional to $(\omega /
\omega_c)^n$ in the spectral function, instead of the $\mathrm{FI}_3$
integral of the ohmic case, one gets
\begin{eqnarray}
\frac{\mbox{FI}_{3+n}}{\omega_c^n} &=& \frac{1}{\omega_c^n}\int_0^\Lambda \frac
{\omega^{3+n} + \mathcal{O}(\omega^{2+n})}{\omega^4 + \mathcal{O}(\omega^3)} d
\omega\\
&=& \frac{1}{\omega_c^n}\int_0^\Lambda \left( \omega^{n-1} + \mathcal{O}\left(\omega^
{n-2}\right) \right) d\omega\\
&=& \frac{1}{\omega_c^n} \left( \frac{\Lambda^n}{n} + \mathcal{O}\left(\Lambda^{n-1}
\right) \right) ,
\end{eqnarray}
and the only way to get a finite non-vanishing contribution is by
taking $\omega_c$ proportional to $\Lambda$ (which was the choice
already made in
Eqs.~\eqref{eq:analytic_spec1}-\eqref{eq:analytic_spec2}).  Therefore,
one just needs to introduce the following simple substitution in the
expressions for the diffusion coefficients of the ohmic case:
\begin{eqnarray}
\mbox{Ohmic} & \to & \mbox{Analytic} \notag\\
\mbox{FI}_3 & \to & \mbox{FI}_3 + \sum_{n=1}^\infty \frac{\gamma_n}{\gamma_0}
\frac{\mbox{FI}_{3+n}}{\Lambda^n}
= \mbox{FI}_3 + \sum_{n=1}^\infty \frac{\gamma_n}{n \gamma_0} ,
\end{eqnarray}
where we took $\omega_c = \Lambda$ and discarded terms involving
negative powers of $\Lambda$. Note that all other integrals
contributing to the diffusion coefficients will exhibit lower powers
of the frequency in the numerator so that their result will be
proportional to inverse powers of $\Lambda$ and can be neglected. The
integrals that correspond to $\mathrm{FC}_3(t)$ and $\mathrm{FS}_4(t)$
in the ohmic case also exhibit divergences that would lead to a
non-vanishing result (and even a divergent one for $\mathrm{FS}_4(t)$)
when dividing by $\Lambda^n$, but that is only for short times of
order $\Lambda^{-1}$ after the initial time. This is the regime where
our explicit expressions are not accurate if the interaction is
gradually switched on with a much longer characteristic time-scale, in
which case the contribution to the diffusion coefficients from that
period is very small.

We close this subsection by pointing out that, since $\mathrm{FI}_3$
is where the $\log{\Lambda}$ divergence arises for the ohmic case, one
could rig an infinite number of finite supraohmic anticoupling terms
to cancel the divergence, though there is no physical motivation to do
so.  One could even ask whether we can both renormalize the diffusion
coefficients and keep the bare frequency finite and unmodified, which
amounts to requiring the following two conditions to hold
simultaneously:
\begin{eqnarray}
\sum_{n=1}^\infty \frac{\gamma_n}{n+1} &=& -\gamma_0 ,\\
\sum_{n=1}^\infty \frac{\gamma_n}{n} &\approx& -\gamma_0 \log{\frac{\Lambda}
{\Omega_r}} .
\end{eqnarray}
The answer is obviously in the negative, at least not with finite
couplings.

\subsection{Laurent Series Spectra Master Equation Coefficients}
\label{sec:master-laurent2}

In this subsection we extend the form of the spectral function
considered in the previous subsection to that of a Laurent
series. This is done by adding an analytic function of $\omega^{-1}$
to the analytic function of $\omega$ already considered there, so that
we have
\begin{equation}
I_{\{\gamma,\varphi\}}(\omega) = I_{\{\gamma\}}(\omega) - \frac{2}{\pi} M \frac
{\lambda}{\omega} \varphi \left( \frac{\lambda}{\omega} \right) \mbox{~~} (\lambda <
\omega)
\label{eq:laurent_spec1},
\end{equation}
where $\varphi$ is an analytic function. Eq.~\eqref{eq:analytic_spec1}
can be written as
\begin{equation}
I_{\{\gamma,\varphi\}}(\omega) = I_{\{\gamma\}}(\omega) - \frac{2}{\pi} M \sum_{n=0}^
\infty \varphi_n \left( \frac{\lambda}{\omega} \right)^{n+1} \mbox{~~} (\lambda <  \omega)
\label{eq:laurent_spec2},
\end{equation}
which corresponds to a linear combination of terms with $N \leq 0$
among those considered in Sec.~\ref{sec:master_general}.


As we saw in Sec.~\ref{sec:master_general}, in the limit $\lambda \to
0$ there is no modification of the equations of motion due to the
terms in Eq.~\eqref{eq:laurent_spec2} with negative powers of
$\omega$. Hence, the dissipation and renormalized frequency terms,
$\Gamma(t),\Omega_{ren}^2(t)$, in the master equation are exactly the
same as with ohmic and, more generally, analytic spectra:
$\Gamma(t)=\gamma_0$, $\Omega_{ren}^2=\Omega_r^2 $. However, there
will be some non-trivial contribution to the diffusion coefficients
that we analyze below.

For times longer than the switch-on time the only difference between
the master equation for the Laurent-series spectrum and the analytic
spectrum lies in a modification of the contribution from the terms
most sensitive to IR divergences, namely $\mathrm{FI}_1$ and
$\mathrm{FC}_1(t)$. This can be seen as follows. When considering a
term proportional to $(\omega_c / \omega)^{n}$ in the spectral
function, instead of the $\mathrm{FC}_1(t)$ integral of the ohmic
case, one gets
\begin{eqnarray}
\omega_c^n \mbox{FC}_{1-n}(t) &=& \omega_c^n \int_\lambda^\infty d \omega
\frac{1}{\omega^n} \frac{(2T / \omega) + \mathcal{O}(\omega^0)}{\Omega_r^4
+ \mathcal{O}(\omega^2)} \cos (\omega t) \nonumber \\
&=& \frac{2T}{\Omega_r^4} \, \omega_c^n \int_\lambda^\infty d \omega
\left[ \omega^{-(n+1)} + \mathcal{O}(\omega^{-(n-1)}) \right] \cos (\omega t) \nonumber \\
&=& \frac{2T}{\Omega_r^4} \, \omega_c^n \left[ \frac{1}{n \lambda^n}
+ \mathcal{O}(\lambda^{-(n-2)}) \right] \cos (\lambda t) ,
\end{eqnarray}
where we used the expansion $\coth(\omega / 2T) = (2T/\omega) +
\mathcal{O} (\omega^0)$ in the first equality. The only way to get a
finite non-vanishing contribution is by taking $\omega_c$ proportional
to $\lambda$ (which was the choice already made in
Eqs.~\eqref{eq:analytic_spec1}-\eqref{eq:analytic_spec2}). While
$\mathrm{FI}_1$ can be obtained by evaluating $\mathrm{FC}_1(t)$ at
$t=0$, all the remaining integrals appearing in the expressions for
the diffusion coefficients will involve a less negative power of the
frequency and, when dividing them by $\lambda^n$, will give a result
corresponding to positive powers of $\lambda$, which can be neglected
in the limit $\lambda \to 0$.

Thus, in order to obtain the diffusion coefficients for a
Laurent-series spectral function of the type considered in this
subsection, one just needs to introduce these simple substitutions in
the expressions for the diffusion coefficients of the ohmic case:
\begin{align}
\mbox{Ohmic} &\to \mbox{Laurent} \notag & \\
\gamma_0 \mbox{FI}_3 &\to \gamma_0 \mbox{FI}_3 + \ell   &   \ell
&= \sum_{n=1}^\infty \frac{\gamma_n}{n} , \\
\gamma_0 \mbox{FC}_1(t) &\to \gamma_0 \mbox{FC}_1(t) - \frac{ 2T }{ \Omega_r^4 }
\cos{(\lambda t)} \phi  &   \phi &= \sum_{n=0}^\infty \frac{\varphi_n}{n+1} , \\
\gamma_0 \mbox{FI}_1 &\to \gamma_0 \mbox{FI}_1 - \frac{ 2T }{ \Omega_r^4 } \phi \, .
& \notag
\end{align}
In summary, the supraohmic terms shift the bare frequency and
$\mathrm{FI}_3$ integrals, and can be used to shift their values in
the ohmic theory, whereas the subohmic terms shift the $\mathrm{FC}_1$
and $\mathrm{FI}_1$ integrals.

\section{Solutions to the Master Equation}
\label{sec:wigner}

\subsection{Solutions to the Laurent-Series Spectra Master Equation}

\subsubsection{Matrix Representation}

First, we express the master equation
\begin{equation}
\frac{\partial}{\partial t} W_r =  \left( -\frac{p}{M} \frac{\partial}{\partial x} + M
\Omega_r^2 x \frac{\partial}{\partial p} + 2 \gamma_0 \frac{\partial}{\partial p} p - 2D_{xp}
\frac{\partial^2}{\partial x \partial p} + D_{pp} \frac{\partial^2}{\partial p^2} \right) W_r \, ,
\end{equation}
in a more compact form:
\begin{equation}
\frac{\partial}{\partial t} W_r = \left( \bolden{\nabla}_\mathbf{q}^{\transpose} \mathbf
{D} \bolden{\nabla}_\mathbf{q} + \bolden{\nabla}_\mathbf{q}^{\transpose} \mathbf{H}
\mathbf{q} \right) W_r \, ,
\end{equation}
with
\begin{align}
\mathbf{q} &= \left[ \begin{array}{c} x \\ p \end{array} \right]    &   \bolden{\nabla}_
\mathbf{q} &= \left[ \begin{array}{c} \frac{\partial}{\partial x} \\ \frac{\partial}{\partial p} \end
{array} \right]\\
\mathbf{H} &= \left[ \begin{array}{cc} 0 & -\frac{1}{M} \\ M \Omega_r^2 & 2\gamma_0
\end{array} \right] &   \mathbf{D} &= \left[ \begin{array}{cc} 0 & -D_{xp} \\ -D_{xp} &
D_{pp} \end{array} \right]
\end{align}
This is a hyperbolic second order partial differential equation
(PDE). As $\mathbf{D}$ is a function of time, in general the equation
is not seperable in time. It is not seperable in phase space either.

\subsubsection{Phase-Space Fourier Transform: The Characteristic Function}

The nature of the PDE suggests a Fourier transform of the phase-space
variables since derivatives are more complicated than algebraic
parameters.  Furthermore, not only does a Fourier transform reduce the
PDE to first order, but the computation of expectation values also
becomes trivial since we are then working with the characteristic
function of the distribution.

The Fourier transform is defined as
\begin{equation}
\mathcal{F}\{f\}(\mathbf{k}) = \int_{-\infty}^{\infty}dx\int_{-\infty}^{\infty}dp \mbox{~} e^
{- \imath \mathbf{k} \cdot \mathbf{q}} f(\mathbf{q}) ,
\end{equation}
and it exhibits the usual properties:
\begin{equation}
\imath^n \frac{\partial^n \mathcal{F}\{f\}}{\partial k_j^n}(\mathbf{0}) = \int_{-\infty}^
{\infty}dx\int_{-\infty}^{\infty}dp \mbox{~} q_j^n f(\mathbf{q}) \label{eq:correlations1}.
\end{equation}
The master equation then becomes
\begin{equation}
\frac{\partial}{\partial t} \mathcal{W}_r = \left( \imath \mathbf{k}^{\transpose} \mathbf
{D} \imath \mathbf{k} + \imath \mathbf{k}^{\transpose} \mathbf{H} \imath \bolden{\nabla}_
\mathbf{k} \right) \mathcal{W}_r ,
\end{equation}
where $\mathcal{W}_r = \mathcal{F}\{W_r\}$ and the normalization of
$W_r (t,\mathbf{q})$ implies $\mathcal{W}_r(t,\mathbf{0}) = 1$.
Finally, it is convenient to group all the derivatives on the
left-hand side, so that we have
\begin{equation}
\left( \frac{\partial}{\partial t} + \mathbf{k}^{\transpose} \mathbf{H}
\bolden{\nabla}_\mathbf{k} \right) \mathcal{W}_r = - \mathbf{k}^{\transpose} \mathbf
{D} \mathbf{k} \mbox{~} \mathcal{W}_r
\label{eq:phase} .
\end{equation}

\subsubsection{Method of Characteristic Curves}

The method of characteristic curves involves looking for parameterized
curves in the domain $(t,\mathbf{k})$ along which the first order PDE
becomes a set of first order ODEs. For each one of those curves we
have
\begin{eqnarray}
\mathcal{W}_r \left(t,\mathbf{k}\right) & = & \mathcal{W}_r \left( t(s), \mathbf{k}(s) \right),\\
\label{eq:cc} \frac{d}{ds} \mathcal{W}_r & = & \frac{dt}{ds} \frac{\partial}{\partial t}
\mathcal{W}_r + \frac{d\mathbf{k}}{ds}^{\transpose} \bolden{\nabla}_\mathbf{k} \mathcal
{W}_r \, ,
\end{eqnarray}
Next, we attempt to match the right-hand side of Eq.~\eqref{eq:cc} to
the left-hand side of Eq.~\eqref{eq:phase}. This results in a system
of ODEs in the parameter $s$. We will look for curves that synchronize
with the initial time so that $t(0)=0$, $\mathbf{k} (0) =
\mathbf{k}_0$. The time solution is simple:
\begin{equation}
\frac{dt}{ds} = 1 \; \Rightarrow \; t(s) = s \, .
\end{equation}
On the other hand, the solution for the Fourier transform of the
phase-space variables is a bit more involved:
\begin{equation}
\frac{d\mathbf{k}}{ds}^{\transpose} = \mathbf{k}^{\transpose}
\mathbf{H} \; \Rightarrow \; \mathbf{k}^{\transpose} = \mathbf{k}_0^{\transpose}
e^{s \mathbf{H}}
\label{eq:ccfps} .
\end{equation}
In order to calculate the exponential of the $\mathbf{H}$ matrix we
diagonalize it, so that
\begin{equation}
\mathbf{k}^{\transpose} \mathbf{H} \mathbf{q}_{\pm} = h_{\pm} \mathbf{k}^
{\transpose} \mathbf{q}_\pm
\label{eq:diagonaliz1} .
\end{equation}
The eigenvalues, eigenvectors, and exponential matrix are given by
\begin{eqnarray}
h_{\pm} &=& \gamma_0 \pm \imath \tilde{\Omega}  \qquad \mathbf{q}_\pm = \left[ \begin{array}{c} 1 \\ -M h_{\pm} \end{array} \right]\label{eq:heigen} ,\\
e^{-t \mathbf{H}} &=& \left( \begin{array}{cc} 2 M \gamma_0 + M (\partial / \partial t) & 1 \\ -M^2 \Omega_r^2 & M (\partial / \partial t) \end{array} \right) G_\mathrm{ret}(t,0)
\label{eq:exponential},
\end{eqnarray}
where $G_\mathrm{ret}(t,0)$ is the retarded Green function in Eq. \eqref{eq:green}.
We now have the rules for transforming back and forth between the
domain coordinates $\left( t, \mathbf{k} \right)$ and the
characteristic curve coordinates $\left( s, \mathbf{k}_0
\right)$. $\mathbf{k}_0$ does not change along the characteristic
curve, but for a given $s$ it uniquely specifies a particular curve
(except at the origin, where all curves intersect).

Using these results, we can immediately apply the method of
characteristic curves to solving Eq.~\eqref{eq:phase} as follows:
\begin{eqnarray}
\frac{d}{ds} \mathcal{W}_r (t(s),\mathbf{k}(s)) & = & -\mathbf{k}^{\transpose} \mathbf{D}(t)
\mathbf{k} \mbox{~} \mathcal{W}_r (t(s),\mathbf{k}(s)) ,\\
\frac{d}{ds} \mathcal{W}_r \left(s,e^{s \mathbf{H}^{\transpose}}\mathbf{k}_0 \right) & = &
-\mathbf{k}_0^{\transpose} e^{s \mathbf{H}} \mathbf{D}(s) e^{s \mathbf{H}^{\transpose}} \mathbf{k}_0
\mbox{~} \mathcal{W}_r (s,e^{s \mathbf{H}^{\transpose}}\mathbf{k}_0) .
\end{eqnarray}
The last equation is a linear ODE whose solution can easily be found
to be
\begin{equation}
\mathcal{W}_r \left(s,e^{s \mathbf{H}^{\transpose}}\mathbf{k}_0 \right) =
\mathcal{W}_r\left( 0,\mathbf{k}_0 \right) e^{-\int_0^s ds' \mathbf{k}_0^{\transpose}
e^{s' \mathbf{H}} \mathbf{D}(s') e^{s' \mathbf{H}^{\transpose}} \mathbf{k}_0} ,
\end{equation}
where $\mathcal{W}_r\left( 0,\mathbf{k}_0 \right)$ is the initial
reduced Wigner function at $t=0$. We can now express the solution back
in terms of $\mathbf{k}$ and get the final result
\begin{equation}
\mathcal{W}_r \left(t, \mathbf{k} \right) = \mathcal{W}_r\left( 0, e^{-t
\mathbf{H}^{\transpose}} \mathbf{k} \right) e^{- \frac{1}{2} \mathbf{k}^{\transpose}
\bolden{\sigma}_T(t) \mathbf{k}}
\label{eq:sol} ,
\end{equation}
with
\begin{equation}
\label{eq:tsig} \bolden{\sigma}_T(t) = 2 \int_0^t e^{(s-t)\mathbf{H}} \mathbf{D}(s) e^
{(s-t)\mathbf{H}^{\transpose}} ds .
\end{equation}
One can check that this result agrees with those in Ref.~\cite{roura}
if one considers equal times in the two-point correlation functions
there.

{From} Eqs.~\eqref{eq:exponential} and \eqref{eq:green} one can see
that all the integrals in Eq.~\eqref{eq:tsig} involve a factor $e^{2
  (s-t) \gamma_0}$ times some oscillatory factor.
Taking into account that
the diffusion coefficients tend to constant asymptotic values $D_{xp}$
and $D_{pp}$ for sufficiently late times, one can see that for $t \gg
\gamma_0$ the integral is dominated by large values of $s$. Therefore,
the asymptotic value of $\bolden{\sigma}_T(t)$ can be calculated using
the constant asymptotic values of the diffusion coefficients in
Eq.~\eqref{eq:tsig} and taking the limit $t \to \infty$, with the
following result:
\begin{equation}
\bolden{\sigma}_T^\infty = \left[ \begin{array}{cc} \frac{1}{(M
\Omega_r)^2} \left( \frac{1}{2 \gamma_0}D_{pp}^\infty - 2 M D_{xp}^\infty \right) & 0 \\
0 & \frac{1}{2 \gamma_0}D_{pp}^\infty \end{array} \right]
\label{eq:ltsig} .
\end{equation}

The solution \eqref{eq:sol} has two factors. The first one tends to
one in the long time limit and encodes the disappearance of the
initial state (we will call it the \emph{death factor}). The second
factor describes the appearance of a Gaussian state that evolves in
time and tends asymptotically to a state that corresponds to thermal
equilibrium (we will refer to this as the \emph{birth factor}).  All
initial distributions evolve towards this final Gaussian state, whose
covariance matrix is given by Eq.~\eqref{eq:ltsig}. This state does
not look like the thermal state of a free harmonic oscillator because
of the coupling to the environment. It results from considering the
thermal equilibrium state for the whole system (system plus
environment) including the system-environment interaction, which
gives rise to a non-trivial correlation between them, and tracing out
the environment.

The \emph{death factor} contains the information on the initial
conditions, describes the gradual disappearance of the initial
distribution and it is always temperature independent.  The initial
distribution undergoes damped oscillations with characteristic
time-scales $\gamma_0, \tilde{\Omega}$.  The higher cumulants of the
distribution, discussed in the next subsection, oscillate and decay
more rapidly. This is responsible for the inspirals in phase space of
the evolution of Gaussians plotted by Unruh and Zurek \cite{unruh},
which are calculated in the next subsection and plotted in
Fig.~\ref{fig:spiral}.
Of course this is all assuming a non-vanishing ohmic term in the
spectral function. If the spectral function is purely non-ohmic
(\emph{i.e.}, with $\gamma_0=0$) then this factor will describe the
initial state oscillating with a renormalized frequency, but which
never decays away.

The \emph{birth factor} describes the complicated birth of and
settlement into a state of thermal equilibrium.  This factor is always
Gaussian with a covariance matrix given by Eq.~\eqref{eq:tsig}, which
involves a convolution of the diffusion matrix with propagators that
reflect the natural oscillatory decay of the system. This covariance
matrix vanishes at the initial time and tends at late times to the
equilibrium covariance matrix \eqref{eq:ltsig}, with the diffusion
coefficients being their asymptotic constant values given by
Eqs.~\eqref{eq:lateDxp}-\eqref{eq:lateDpp}.  This covariance matrix is
positive definite for all reasonable diffusion constants.  Moreover,
the anomalous diffusion coefficient actually acts as an
``anti-diffusion'' term that makes $\sigma_{xx}$ (and the uncertainty
in $x$) free of the $\log (\Lambda / \Omega_r)$ divergence, as it will
be discussed in the next subsection.

\subsection{Analysis of the Solutions}
\subsubsection{Trajectories of the Cumulants}
\label{sec:wigner-sol}

\begin{figure}[h]
\centering
\includegraphics[width=4in]{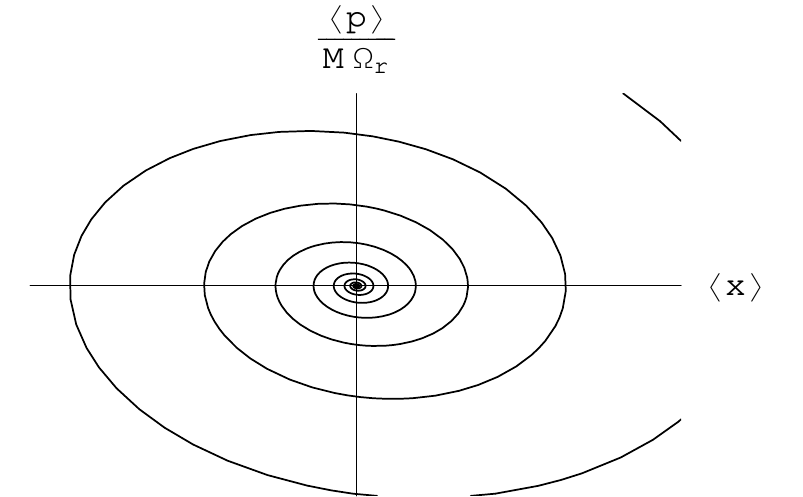}
\caption{\label{fig:spiral} The trajectory of the expectation values
$\langle x \rangle, \langle p \rangle$.}
\end{figure}

As we have already mentioned, the Fourier transfom of the reduced
Wigner function corresponds to its characteristic function, from which
the correlation functions for the phase-space variables can be easily
derived using Eq.~\eqref{eq:correlations1}. The general expressions
for the cumulants can be obtained straightforwardly from the logarithm
of the reduced Wigner function in Fourier space as follows:
\begin{equation}
\sum_{n=1}^{\infty} \frac{1}{n!} \kappa^{(n)}_{i_1 \ldots i_n}(t)
\prod_{l=1}^{n} \imath k^{i_l}
= \log{\mathcal{W}_r(t,\bolden{\kappa})} ,
\end{equation}
where $k^{i_l}$ denotes the components of the vector $\mathbf{k}$ and
we used the Einstein summation convention for pairs of repeated
indices (\emph{i.e.}, it is implicitly understood that a sum
$\sum_{i_l = 1}^{2}$ should be preformed over each pair of repeated
indices $i_l$). $\bolden{\kappa}^{(n)}$ is the $n^{th}$ cumulant and
acts as a tensor of order $n$ contracted with $n$ copies of
$\mathbf{k}$. Using the result for $\mathcal{W}_r(t,\bolden{\kappa})$
from Eq.~\eqref{eq:sol} we have
\begin{equation}
\sum_{n=1}^{\infty} \frac{1}{n!} \kappa^{(n)}_{i_1 \ldots i_n}(t)
\prod_{l=1}^{n} \imath k^{i_l} =
\sum_{n=1}^{\infty} \frac{1}{n!} \kappa^{(n)}_{i_1 \ldots i_n}(0)
\prod_{l=1}^{n} \imath \left( e^{-t \mathbf{H}^{\transpose}}\mathbf{k}\right)^{i_l}
- \frac{1}{2} \mathbf{k}^{\transpose} \bolden{\sigma}_T(t) \mathbf{k}
\label{eq:cumulants2} ,
\end{equation}
where $\kappa^{(n)}_{j_1 \ldots j_n} (0)$ are the cumulants associated
with the initial distribution. Eq.~\eqref{eq:cumulants2} implies
\begin{equation}
\kappa^{(n)}_{i_1 \ldots i_n}(t) = \kappa^{(n)}_{j_1 \ldots j_n} (0)
\prod_{l=1}^{n} \imath \left( e^{-t \mathbf{H}^{\transpose}} \right)^{j_l i_l}
+ \delta_{n2} \, \sigma_T^{i_1 i_2} (t) .
\end{equation}
We can see that the only cumulant with a non-vanishing asymptotic
value, which is a consequence of the thermal fluctuations, is the
covariance matrix (with $n=2$). The closely related second momenta of
the distribution are given by
\begin{equation}
\langle \mathbf{q} \mathbf{q}^{\transpose} \rangle(t) = e^{-t\mathbf{H}^{\transpose}}
\langle \mathbf{q} \mathbf{q}^{\transpose} \rangle_0 e^{-t\mathbf{H}}
+ \bolden{\sigma}_T(t)
\label{eq:covariance2} ,
\end{equation}
where $\langle \cdots \rangle_0$ denotes expectation value with
respect to the reduced Wigner function at the initial time.  All other
cumulants experience oscillatory decay with time scales of $n
\gamma_0,\, n \tilde{\Omega}$ with $n$ being the order of the
cumulant. In particular, the expectation value
\begin{equation}
\langle \mathbf{q} \rangle(t) = e^{-t\mathbf{H}^{\transpose}} \langle \mathbf{q}
\rangle_0 \, ,
\end{equation}
follows a trajectory plotted in Fig.~\ref{fig:spiral}, where one can
see that the trajectory of the expectation values $\langle x \rangle,
\langle p \rangle$ for any initial distribution inspiral into the
origin.

\subsubsection{Late Time Uncertainty Function}
\label{sec:uncertainty}

\begin{figure}[h]
\centering
\includegraphics[width=4in]{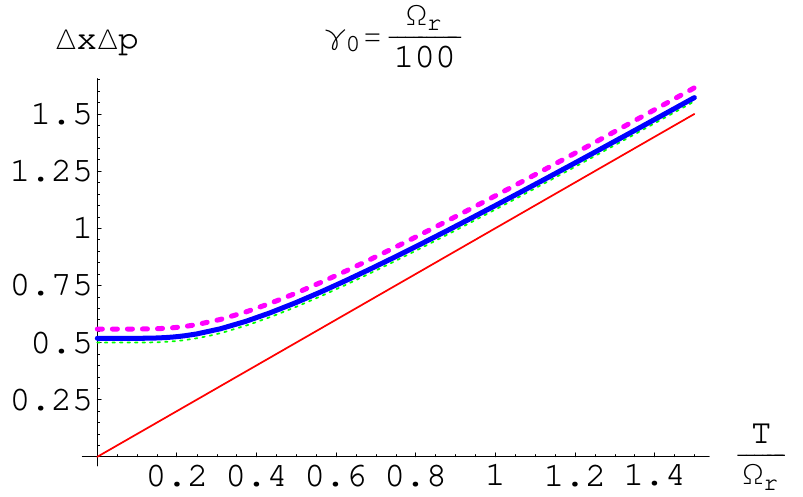}
\caption{\label{fig:dxdp}Late time $\Delta x \Delta p$ for \textcolor{red}{$-$ high
temperature, classical statistical mechanics}, \textcolor{green}{$\cdots$ weak coupling
approximation $\frac{1}{2}\coth{\frac{\Omega_r}{2T}}$}, \textcolor{blue}{\boldmath$-$
\textbf{HPZ at} $\Lambda=10^3 \Omega_r$}, and \textcolor{magenta}{\boldmath$\cdots$
\textbf{HPZ at} $\Lambda=10^9 \Omega_r$}.}
\end{figure}

As we have seen above, any specific features of the initial
distribution decay away (assuming a non-vanishing ohmic term in the
spectral function) and at late times the state tends generically to a
Gaussian with a covariance matrix given by
Eq.~\eqref{eq:ltsig}. Therefore, from Eq.~\eqref{eq:covariance2} it
follows that at late times $(\Delta x)^2 =
(\bolden{\sigma}^\infty_T)_{xx}$ and $(\Delta p)^2 =
(\bolden{\sigma}^\infty_T)_{pp}$. From Eq.~\eqref{eq:ltsig} and using
Eqs.~\eqref{eq:Dxp2}-\eqref{eq:Dpp2} we can express the position and
momentum uncertainties at late times as
\begin{eqnarray}
(\Delta x)^2 &=& \frac{2}{\pi} \frac{\gamma_0}{M} \mbox{FI}_1 ,\\
(\Delta p)^2 &=& \frac{2}{\pi} \gamma_0 M \mbox{FI}_3 .
\end{eqnarray}
The product of the two uncertainties
\begin{equation}
(\Delta x)^2 (\Delta p)^2 = \left( \frac{2\gamma_0}{\pi} \right)^2 \mbox{FI}_1
\mbox{FI}_3 ,
\end{equation}
can be expanded for high temperatures as
\begin{equation}
(\Delta x)^2 (\Delta p)^2 = \left(\frac{T}{\Omega_r}\right)^2 + (\cdots)T + (\cdots)T^0 .
\end{equation}
Inspecting the terms in powers of $T$ immediately reveals the
high-temperature result of classical statistical mechanics for the
case of an ohmic spectrum.  Integer subohmic terms would strictly
decrease this amount.  One can also see that at weak coupling the
uncertainty function agrees with the weak coupling approximation for
moderate values of the cut-off scale, as shown in Fig.~\ref{fig:dxdp}.

\begin{figure}[h]
\centering
\includegraphics[width=4in]{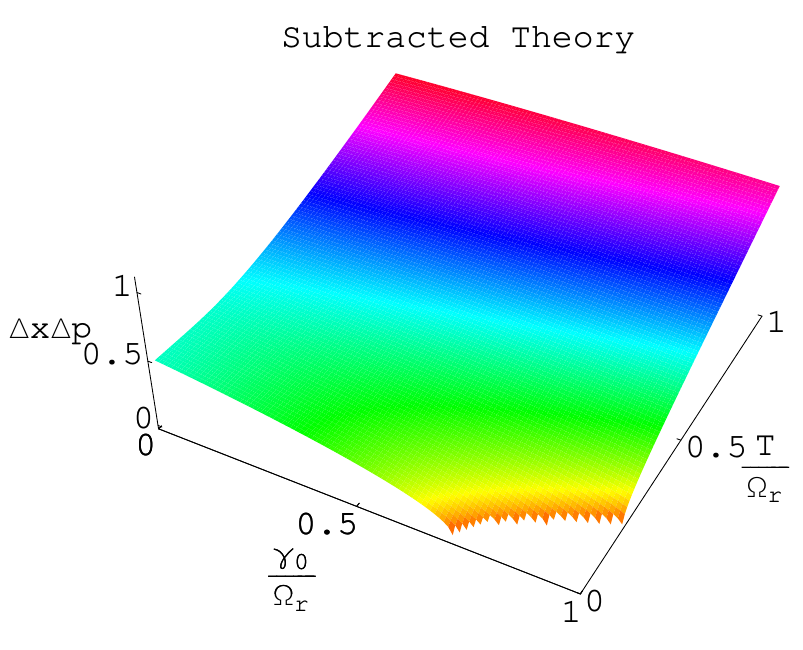}
\caption{\label{fig:dxdp3d-ren}Late time $\Delta x \Delta p$ for the subtracted theory.}
\end{figure}

\begin{figure}[h]
\centering
\includegraphics[width=4in]{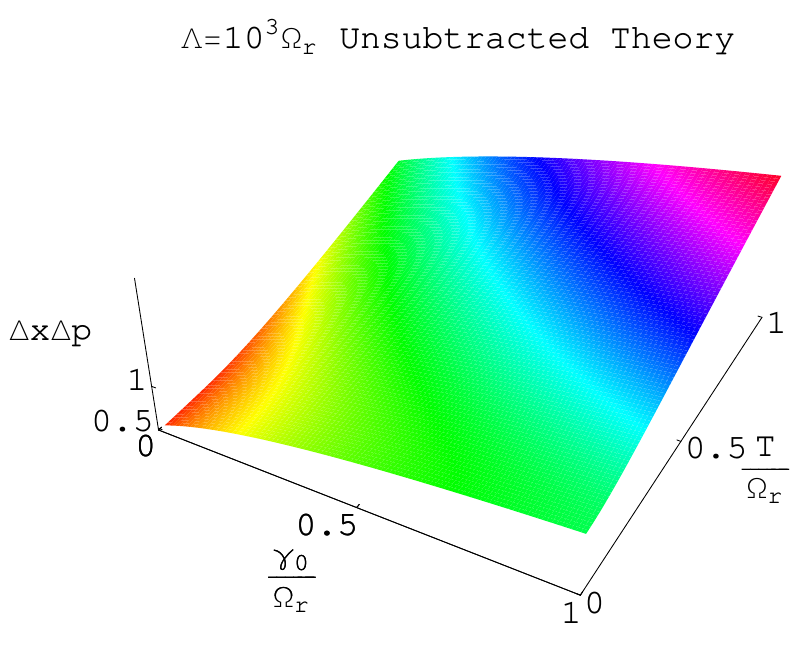}
\caption{\label{fig:dxdp3d-unren}Late time $\Delta x \Delta p$ for the unsubtracted theory.}
\end{figure}

Had one naively tried to have finite diffusion coefficients in the
limit $\Lambda \rightarrow \infty$ subtracting by hand the
$\log(\Lambda / \Omega_r)$ term, one would find a violation of the
Heisenberg uncertainty principle at low temperature and strong
coupling (see Fig.\ref{fig:dxdp3d-ren}), which renders the theory
unphysical. Of course this does not happen with the unsubtracted
theory, as seen in Fig.~\ref{fig:dxdp3d-unren}. It is thus clear that
the logarithmic dependence on the ultraviolet cut-off that appears in
the diffusion coefficients is a physically important parameter and not
something that can be subtracted away.

{From} Eqs.~\eqref{eq:FI_1}-\eqref{eq:FI_3} [or alternatively from
Eq.~\eqref{eq:FI_N}] one can see that mentioned earlier, only the
momentum uncertainty contains a logarithmically divergent cut-off
dependence.  In contrast, the position uncertainty is much smaller and
finite in the limit $\Lambda \rightarrow \infty$ (this had already
been noticed for Gaussian wave-packets in Ref.~\cite{unruh}).  In
fact, if the Brownian particle is coupled to a reservoir at
arbitrarily low temperature and arbitrarily strong
coupling,\footnote{In this limit one needs to use the expressions for
  the overdamping regime, as explained in
  footnote~\ref{overdamping}. Moreover, one needs to keep $\Lambda >
  \gamma_0$, which means that $\Lambda$ also tends to infinity as
  $\gamma \to \infty$. Alternatively, one should use
  Eqs.~\eqref{eq:FI_1}-\eqref{eq:FI_3} plus the terms involving
  negative powers of $\Lambda$ which were neglected there, so that the
  restriction $\Lambda > \gamma_0$ does not apply. In both cases one
  gets the result quoted in Eqs.~\eqref{eq:deltaX}-\eqref{eq:deltaP}.}
the uncertainty in position will be arbitrarily small:
\begin{eqnarray}
\lim_{\gamma_0 \to \infty} \lim_{T \to 0} \Delta x &=& 0 , \label{eq:deltaX}\\
\lim_{\gamma_0 \to \infty} \lim_{T \to 0} \Delta p &=& \infty \label{eq:deltaP}.
\end{eqnarray}
Hence, strong coupling to a low temperature reservoir tends to
localize the Wigner distribution in position. From this result, one
would qualitatively expect strong coupling to suppress quantum
tunneling while high temperature would aid both tunneling and
classical escape (thermal activation).

\subsubsection{Linear Entropy}
\label{sec:entropy}

In this subsection we investigate the linear entropy~\cite{feix},
which can be easily obtained from the Wigner distribution as follows:
\begin{equation}
S_\mathrm{L} = 1 - \mbox{Tr}(\hat{\mathbf{\rho}}^2_r)
= 1 - 2\pi \int d^2\mathbf{q} W_r^2 (\mathbf{q}) .
\end{equation}
In Fourier space it becomes
\begin{equation}
S_L = 1 - \frac{1}{2\pi} \int d^2\mathbf{k} |\mathcal{W}_r(\mathbf{k})|^2 ,
\end{equation}
and using the result in Eq.~\eqref{eq:sol} we finally get
\begin{equation}
S_L = 1 - \frac{1}{2\pi} \int d^2\mathbf{k}
\left|\mathcal{W}_r \left( 0, e^{-t\mathbf{H}^{\transpose}} \mathbf{k} \right) \right|^2
e^{- \mathbf{k}^{\transpose} \bolden{\sigma}_T(t) \mathbf{k}}
\label{eq:lin-ent1} .
\end{equation}
At the initial time the linear entropy is that of the initial state,
and at late times it tends to $S_L = 1 - (1/2)(\det
\bolden{\sigma}_T^\infty)^{-1/2}$.

Alternatively, one can express the linear entropy in terms of an
integral of the Fourier-transformed reduced Wigner function at the
initial time by introducing the change of variables $\mathbf{k_0} =
e^{-t\mathbf{H}^{\transpose}} \mathbf{k}$.  Eq.~\eqref{eq:lin-ent1}
can then be written as
\begin{eqnarray}
S_L &=& 1 - \frac{1}{2\pi} \int d^2 \mathbf{k_0} \det{\left( e^{+t\mathbf{H}^{\transpose}} \right)}
|\mathcal{W}_r\left( 0, \mathbf{k_0} \right)|^2 e^{- \mathbf{k_0}^{\transpose} e^{+t\mathbf{H}}
\bolden{\sigma}_T(t) e^{+t\mathbf{H}^{\transpose}} \mathbf{k_0}}
\nonumber \\
&=& 1 - \frac{1}{2 \det^{1/2} [\bolden{\sigma}_T(t)]} \int d^2\mathbf{k_0}
|\mathcal{W}_r\left( 0, \mathbf{k_0} \right)|^2 N\left( \mathbf{0}, e^{+t\mathbf{H}}
\bolden{\sigma}_T(t) e^{+t \mathbf{H}^{\transpose}}; \mathbf{k_0} \right) ,
\end{eqnarray}
where $N(\bolden{\mu},\bolden{\sigma}; \mathbf{k_0})$ is a normalized
Gaussian distribution for the variable $\mathbf{k_0}$ with mean
$\bolden{\mu}$ and covariance $\bolden{\sigma}$.  For small times this
integral is similar to that for the initial state, whereas for long
times the normalized Gaussian distribution becomes increasingly close
to a delta function.

For a Gaussian initial state $\mathcal{W}_r (0, \mathbf{k_0}) =
\exp[- \mathbf{k_0}^{\transpose} \bolden{\sigma}_T(t) \mathbf{k_0}
- \imath \mathbf{k_0} \langle \mathbf{q} \rangle_0]$ the integral in
Eq.~\eqref{eq:lin-ent1} can be explicitly computed:
\begin{eqnarray}
S_L &=& 1 - \frac{1}{2\pi} \int d^2\mathbf{k} e^{- \mathbf{k}^{\transpose} \left( e^{-t
\mathbf{H}} \bolden{\sigma}_0 e^{-t\mathbf{H}^{\transpose}} + \bolden{\sigma}_T(t)
\right) \mathbf{k}} \nonumber \\
&=& 1 - \frac{1}{2 \det^{1/2} \! \left[ e^{-t\mathbf{H}} \bolden{\sigma}_0
e^{-t\mathbf{H}^{\transpose}} + \bolden{\sigma}_T(t) \right]} .
\end{eqnarray}
For these Gaussian states, reasonable linear entropy is synonymous
with reasonable uncertainty functions (\emph{i.e.}, the linear entropy
will be positive if and only if the Heisenberg uncertainty principle
is satisfied).  We have already found that the late time uncertainty
is well behaved. The uncertainty at the initial and intermediate times
should not violate the Heisenberg uncertainty principle either.  As a
particular example, let us consider an initial state that corresponds
to the ground state associated with the bare frequency. At the initial
time the determinant is completely specified by the determinant of $
\bolden{\sigma}_0$, which equals $1/4$ as a consequence of the
cancellation between the large $(\Delta p)^2$ factor (of the order of
$M \Omega_\mathrm{bare}$ with $\Omega_\mathrm{bare} \sim
\sqrt{\Lambda}$) and the small $(\Delta x)^2$ factor (of order $1 / M
\Omega_\mathrm{bare}$). However, after a very short time, when
$\bolden{\sigma}_T(t)$ starts acquiring non-vanishing values, it will
dominate the total contribution to $(\Delta x)^2$ and the linear
entropy will increase suddenly (this kind of behavior for the entropy
was found in Ref.~\cite{lin}).
At later times it will relax to its thermal equilibrium values.

\subsection{Solutions of the General Master Equation}
\label{sec:wigner-gen}

In situations more general than the Laurent series spectrum and where
the classical equation of motion involves fractional calculus
(\emph{i.e.}, it becomes an integro-differential equation rather than
an ODE), the master equation can be of a slightly more general nature,
with additional time dependence in the frequency and the dissipation
coefficients:
\begin{equation}
\frac{\partial}{\partial t} W_r = \left( -\frac{p}{M} \frac{\partial}{\partial x} + M
\Omega_r^2(t) x \frac{\partial}{\partial p} + 2 \Gamma(t) \frac{\partial}{\partial p} p +
\bolden{\nabla}_\mathbf{q}^{\transpose} \mathbf{D}(t) \bolden{\nabla}_\mathbf{q} \right)
W_r \, .
\end{equation}
The method for finding its solutions will be almost the same, except
that solving for the characteristic curves of the Fourier-transformed
phase-space variables is now less straightforward due to the time
dependence of $\mathbf{H}$, and the procedure followed to solve
Eq.~\eqref{eq:ccfps} needs to be generalized.  The system of ODEs that
corresponds to
\begin{eqnarray}
\frac{d\mathbf{k}}{ds}^{\transpose} &=& \mathbf{k}^{\transpose} \mathbf{H}(s) ,
\end{eqnarray}
can be decoupled through a process of differentiation and substitution
that leads to the equations for a pair of independent parametric
oscillators where solving one effectively solves the other up to an
integral. The second order ODE satisfied by $k_p$ is
\begin{equation}
0 = \ddot{k}_p - 2 \Gamma \dot{k}_p + \left( \Omega_r^2 - 2\dot{\Gamma} \right) k_p \, .
\end{equation}
After solving it, $k_x$ can be simply obtained as
\begin{equation}
k_x = k_x^0 + M \int_0^s ds' \mbox{~} \Omega_r^2(s') k_p(s') .
\end{equation}
Alternatively, one can proceed in the reverse order as follows:
\begin{eqnarray}
0 &=& \ddot{k}_x - \left( 2 \Gamma + \frac{\dot{\Omega}_r^2}{\Omega_r^2} \right) \dot{k}
_x + \Omega_r^2 k_x \, , \\
k_p &=& \left( k_p^0 - \frac{1}{M} \int_0^s ds' \mbox{~} e^{-2\int_0^{s'} \Gamma(s'') ds''} ,
k_x(s') \right) e^{+2\int_0^s \Gamma(s') ds'} .
\end{eqnarray}
Solving for $k_p$ first appears to be simpler since everything is then
linear in the coefficients. It is convenient to factor out the
exponential growth, which reduces the ODE to its undamped form:
\begin{eqnarray}
k_p & = & e^{\int_0^s \Gamma(s')ds'} j_p \, ,\\
0 &=& \ddot{j}_p + \left( \Omega_r^2 -\Gamma^2 - \dot{\Gamma} \right) j_p \, .
\end{eqnarray}
If one is able to solve this differential equation, which for simple
enough functions could be performed with Floquet analysis and
variation of parameters, then the characteristic curves can be
expressed in terms of the following matrix equation:
\begin{equation}
\mathbf{k} = \bolden{\Phi}(s) \mathbf{k}_0 \, ,
\end{equation}
where $\bolden{\Phi}(s)$ is the so-called transition matrix, which has
many of the properties of an exponential and contains the parametric
oscillatory behavior of the characteristic curves.
If $\Gamma(t)$ and $\Omega_r(t)$ tend to some asymptotic values at
large times, then the behavior of $\bolden{\Phi}(s)$ tends to that of
a damped harmonic oscillator.

Once the transition matrix is available one can apply exactly the same
approach as above in order to solve the master equation. One starts by
writing the master equation as
\begin{equation}
\frac{d}{ds} \mathcal{W}_r (t(s),\bolden{\Phi}(s) \mathbf{k}_0)
= - \mathbf{k}_0^{\transpose} \bolden{\Phi}^{\transpose}(s) \mathbf{D} (s)
\bolden{\Phi}(s) \mathbf{k}_0 \mbox{~} \mathcal{W}_r (t(s),\bolden{\Phi}(s) \mathbf{k}_0).
\end{equation}
Next, one solves this linear ODE in $s$ and, reexpressing the result
in terms $t$ and $\mathbf{k}$, one can finally write the solution of
the master equation as
\begin{equation}
\mathcal{W}_r \left(t, \mathbf{k} \right)
= \mathcal{W}_r \left( 0, \bolden{\Phi}^{-1}(t) \mathbf{k} \right)
e^{- \frac{1}{2} \mathbf{k}^{\transpose} \bolden{\sigma}_T(t) \mathbf{k}}
\label{eq:solg},
\end{equation}
with
\begin{equation}
\bolden{\sigma}_T(t) = 2 \left( \int_0^t [\bolden{\Phi}^{-1}(t)]^{\transpose}
\bolden{\Phi}^{\transpose} (s) \mathbf{D}(s) \bolden{\Phi}(s) \bolden{\Phi}^{-1}(t) ds
\right) .
\end{equation}

For ohmic-like coefficients, the qualitative behavior of the solution
\eqref{eq:solg} is very similar to that of the ohmic solution. The
moments of the initial distribution will experience oscillatory decay
and the solution will tend to a thermal equilibrium state at late
times.  In purely supraohmic regimes where $\Gamma$ vanishes, the
initial state would remain with renormalized frequency.  In purely
subohmic regimes where $\Gamma$ grows quickly, the initial state would
vanish quickly.

%
\section{Influence of a Classical Force}
\label{sec:force}

In this section we consider the case of a classical force $F(t)$
acting on the quantum oscillator. This is done by introducing a
time-dependent potential $-F(t)x$:
\begin{equation}
L_x = \frac{1}{2}M \left( \dot{x}^2 - \Omega^2 x^2 \right) + F(t) x .
\end{equation}

\subsection{The Master Equation Coefficients}

To derive the form of the master equation we follow the method of
Calzetta, Roura, and Verdaguer \cite{roura} as it is of a very general
nature and can be adapted quite straightforwardly.  In the presence of
an external force the system action takes the following form in terms
of the ``center of mass'' and ``relative'' coordinates, $X=(x+x')/2$
and $\Delta=x-x'$:
\begin{eqnarray}
S(x)-S(x') &=& M\int_0^t \left( M ds \left[ \dot{X}(s) \dot{\Delta}(s) - \Omega^2 X(s) \Delta
(s) \right] + F(s) \Delta(s) \right) .
\end{eqnarray}
The reduced Wigner function can still be expressed as an average over
a stochastic process and a distribution for the initial conditions as
follows:
\begin{eqnarray}
W_r(x,p,t) &=& \left\langle \left\langle \delta\left(X(t)-x\right)
\delta\left(M\dot{X}(t)-p\right),
\right\rangle_\xi \right\rangle_{X_0,p_0} \, , \\
\langle \dots \rangle_\xi & \equiv & \int \mathcal{D}\xi \dots
e^{-\frac{1}{2}\xi \cdot \nu^{-1} \cdot \xi} \, ,
\end{eqnarray}
where we used the notation $\cdot \equiv \int_0^t$ in the last
equation, and $X(t)$ is now a solution to the Langevin equation with
the external force:
\begin{eqnarray}
M \left( \frac{d^2}{dt^2} + 2\gamma_0 \frac{d}{dt} + \Omega_r^2 \right) X(t) &=& F(t) + \xi
(t) .
\label{eq:langevin1}
\end{eqnarray}
Differentiating with respect to time reveals the usual dissipation,
renormalized harmonic potential and the classical force potential
$-F(t)x$:
\begin{eqnarray}
\frac{\partial}{\partial t} W_r &=& \left( \bolden{\nabla}_\mathbf{q}^{\transpose} \mathbf{H}
\mathbf{q} - F(t) \frac{\partial}{\partial p} \right) W_r - \frac{\partial}{\partial p}
\left\langle
\left\langle \xi(t) \delta\left(X(t)-x\right) \delta\left(M\dot{X}(t)-p\right) \right\rangle_{\xi}
\right\rangle_{X_0,p_0} \, .
\end{eqnarray}
After a functional integration by parts, the last term gives rise to
the diffusion terms, with the following coefficients:
\begin{eqnarray}
D_{xp} &=& -\frac{1}{2} \int_0^t ds \mbox{~} \nu(t,s) \frac{\delta X(t)}{\delta \xi(s)} , \\
D_{pp} &=& M \int_0^t ds \mbox{~} \nu(t,s) \frac{\partial}{\partial t} \frac{\delta X(t)}
{\delta \xi(s)} .
\end{eqnarray}
Eq.~(\ref{eq:langevin1}) can be solved in the same way as in the case
without external force to obtain $X(t)$. It is simply given by a
homogeneous solution that contains all the information on the initial
conditions, plus a convolution of the external force with the retarded
propagator associated with the homogeneous part of the equation, plus
a convolution of the stochastic source with the same retarded
propagator.  By the linearity of the functional derivative,
$\frac{\delta X(t)}{\delta \xi(s)}$ only depends on the retarded
propagator, which is the same as in the case of no external force.
Therefore, the master equation has exactly the same form, except for
the addition of the classical force in the potential derivative term
of the Poisson bracket:
\begin{eqnarray}
\frac{\partial}{\partial t} W_r & = & \left( \bolden{\nabla}_\mathbf{q}^{\transpose} \mathbf
{D} \bolden{\nabla}_\mathbf{q} + \bolden{\nabla}_\mathbf{q}^{\transpose} \mathbf{H}
\mathbf{q} - F(t) \frac{\partial}{\partial p} \right) W_r \, .
\end{eqnarray}

\subsection{Solutions of the Master Equation}
\label{sec:force-sol}

Fourier transforming the phase-space variables, the master equation becomes
\begin{eqnarray}
\left( \frac{\partial}{\partial t} + \mathbf{k}^{\transpose} \mathbf{H} \bolden{\nabla}_
\mathbf{k} \right) \mathcal{W}_r & = & -\left( \mathbf{k}^{\transpose} \mathbf{D} \mathbf
{k} + \imath F(t) k_p \right) \mathcal{W}_r \, .
\end{eqnarray}
This equation can also be solved via the method of characteristic
curves, following the same method of the previous section:
\begin{eqnarray}
\frac{d}{ds} \mathcal{W}_r (t(s),\mathbf{k}(s)) & = &
-\left( \mathbf{k}^{\transpose} \mathbf{D} \mathbf{k} +
\imath F(s) k_p \right) \mathcal{W}_r (t(s),\mathbf{k}(s)) , \\
\frac{d}{ds} \mathcal{W}_r \left(s,e^{s \mathbf{H}^{\transpose}}\mathbf{k}_0 \right) & = &
-\left( \mathbf{k}_0^{\transpose} e^{s \mathbf{H}}
\mathbf{D}(s) e^{s \mathbf{H}^{\transpose}} \mathbf{k}_0 + \imath F(s) \hat{\mathbf{k}}_
{p}^{\transpose} e^{s \mathbf{H}^{\transpose}} \mathbf{k}_0 \right) \mathcal{W}_r
\left(s,e^{s \mathbf{H}^{\transpose}}\mathbf{k}_0 \right) ,
\end{eqnarray}
with the solution
\begin{equation}
\mathcal{W}_r \left(s, e^{s \mathbf{H}^{\transpose}} \mathbf{k}_0 \right)
= \mathcal{W}_r\left( 0,\mathbf{k}_0 \right)
e^{-\int_0^s ds' \left( \mathbf{k}_0^{\transpose} e^{s' \mathbf{H}} \mathbf{D}(s') e^{s'
\mathbf{H}^{\transpose}} \mathbf{k}_0 + \imath F(s') \hat{\mathbf{k}}_{p}^{\transpose} e^
{s' \mathbf{H}^{\transpose}} \mathbf{k}_0 \right)} ,
\end{equation}
which can be finally written as
\begin{equation}
\mathcal{W}_r \left(t, \mathbf{k} \right) =  \mathcal{W}_r\left( 0, e^{-t\mathbf{H}^
{\transpose}} \mathbf{k} \right) e^{- \frac{1}{2} \mathbf{k}^{\transpose} \bolden{\sigma}_T
(t) \mathbf{k} - \imath \left(\langle \mathbf{k} \rangle_F(t)\right)^{\transpose} \mathbf{k}}
\, ,
\end{equation}
where
\begin{equation}
\langle \mathbf{k} \rangle_F(t) = \int_0^t ds \mbox{~} F(s) e^{(s-t)
\mathbf{H}} \hat{\mathbf{k}}_{p}
\label{eq:response} \, .
\end{equation}
One can see that just as all temperature dependence only appears in the
second cumulant, or covariance, the external force only affects the first
cumulant, or mean.  The change in the evolution of the mean, given by
Eq.~\eqref{eq:response}, simply corresponds to adding the response to
the external force.  As with an ordinary driven oscillator, the
response is a convolution of the retarded propagator, which exhibits
the natural oscillatory decay of the system, and the driving force.

\section{Summary of Results}

Quantum Brownian motion of an oscillator coupled to a thermal
reservoir of quantum oscillators has been the canonical model for
studying the environmental effects on a quantum system, even of
macroscopc scale, such as quantum dissipation, diffusion, decoherence
and entanglement. It also provides important information on quantum
measurement, such as noise, fluctuations, correlations, uncertainty
relation and standard quantum limit in mesoscopic systems.  Many
experiments have been carried out for testing these processes.
Fifteen years ago an exact master equation \cite{QBM1} for the
reduced density matrix of the system were derived for a general
environment of arbitrary spectral density and temperature. There are
claims that exact solutions have been found \cite{solnHPZ}. In this
paper we report on solutions to this equation for a fairly general
set of physical conditions and analyze its salient features. We
expect these solutions to be useful in realistic settings for the
analysis of many problems which can be described by this model.

The first question we addressed concerns the initial time
divergences. These divergences result from the sudden coupling of
an initially uncorrelated system and environment.
This situation is unphysical and can be resolved by switching on the
system-environment interaction smoothly over a time-scale longer than
the inverse of the characteristic UV cut-off scale of the environment.
Having switched on the coupling, the initial frequency of the
oscillator, as it quickly evolves from the bare to the renormalized
frequency, is the only remaining initial singular behavior.
But this merely implies a transition from the bare initial state to a
``renormalized'' version of that state, which can be taken as the
effective initial state for the subsequent regular evolution.

We then solved the master equation coefficients for a general class
of bilinear system-environment couplings, which includes any
combination of ohmic and integer supraohmic and subohmic spectral
densities with extreme cut-offs.  Although it is by no means a
complete class of spectral densities, it is a privileged class in
that the classical trajectories are determined by ordinary
differential equations and not integro-differential or fractional
differential equations.  In that sense, their dynamics is more
classical.

For these system-environment couplings, we have solved the master
equation coefficients for all temperature ranges, beyond weak
coupling, and at all times after the short initial time-scale during
which the coupling is switched on. This has enabled us to clarify the
validity and shortcomings of previously obtained approximations and
provide more general results.  Perhaps the most useful are our simple
expressions for the master equation coefficients at late time
[Eq.~\eqref{eq:lateDxp}-\eqref{eq:lateDpp}]. They clearly reveal the
existence of logarithmic divergences in the limit of infinite UV
cut-off not only in the anomalous diffusion coefficient, but also in
the normal one. These divergences cannot be consistently subtracted
and condition the possibility of obtaining meaningful results upon
the existence of a physically well-motivated cut-off.

For the system-environment couplings under discussion, which all have
local dissipation, we have also solved the master equation in its
entirety.  All initial states evolve into a given Gaussian state
corresponding to thermal equilibrium.  We have obtained the
covariance matrix for this state [Eq.~\eqref{eq:ltsig}], and the
corresponding uncertainty function (Sec.~\ref{sec:uncertainty}).
Interestingly, arbitrarily strong coupling to a zero temperature
reservoir will tend to localize the position with arbitrary
precision.

For more general systems with non-local dissipation, we have
drastically reduced the task of solving the master equation to that
of solving a one-dimensional classical parametric oscillator problem.
The master equation solutions are parametrically similar to the case
of local dissipation. The same kind of terms arise, including a
thermal covariance, but it is not guaranteed that the system will
relax to a thermal state.  That of course depends upon the behavior
of the classical trajectories.

Finally, we extended the model of the quantum oscillator linearly
coupled to a thermal reservoir of oscillators by including a classical
driving force. This modifies the dynamics by driving the mean position
and momentum around just as with a classical driven system.  In this
model we found that the force has no effect upon the width of the
wave-packet or any cumulant other than the mean.  These results may be
useful for the study of low-temperature measurements of forced
oscillators, which are relevant for experiments with nanomechanical
resonators \cite{naik,lahaye}. They also play a crucial role in future
schemes for the detection of gravitational waves with high-intensity
laser interferometers, where the radiation pressure effects on the
cavity mirrors are important \cite{kimble,buonanno}.

\acknowledgments

C.~F. and B.~L.~H. are  supported in part by grants from the NSF-ITR
program (PHY-0426696), NIST and NSA-LPS. A.~R.\  is supported by
LDRD funds from Los Alamos National Laboratory.

\appendix
\section*{Appendix}
\setcounter{section}{1}


\subsection{Harmonic Number}
\label{sec:harmonic}

The Harmonic Number $H(n)$ is a function similar to a logarithm. Its
analytical continuation to the complex plane is similar too.
\begin{eqnarray}
\mbox{H}(n) & = & \sum_{k=1}^n \frac{1}{k}\, , \qquad n \in \mathbb{Z}^+ \\
\mbox{H}(0) & = & 0 , \\
\gamma_E & = & \lim_{n \rightarrow \infty} (\mbox{H}(n)-\log{(n)}) , \\
\mbox{H}(z) & = & \gamma_E + \psi(z+1) ,  \qquad z \in \mathbb{C} \\
\psi(z) & = & \frac{\Gamma'(z)}{\Gamma(z)} , \\
\psi(z) & \sim & \ln z -\frac{1}{2 z} - \frac{1}{12 z^2} + \cdots  \mbox{~ if ~}
\left| \arg{(z)} \right| < \pi
\label{psi2} ,
\end{eqnarray}
where Eq.~(\ref{psi2}) is the asymptotic expansion for $|z| \to \infty$.

\subsection{Exponential Integrals}
\label{sec:expint}

The following definition and properties of the exponential integral
are used in this paper:
\begin{eqnarray}
\mbox{Ei}(x) & = & -\, \mathrm{P.V.} \int_{-x}^\infty dt \, \frac{e^{-t}}{t}  \mbox{~ if ~} x > 0
\label{ei1}, \\
\mbox{E}_1(z) & = & \int_z^\infty dt \, \frac{e^{-t}}{t} \mbox{~ if ~} \left| \arg{(z)} \right|
< \pi , \\
\mbox{E}_1(z) & \sim & \frac{e^{-z}}{z} \left( 1 - \frac{1}{z} + \frac{2}{z^2} + \cdots
\right) \label{ei3},
\end{eqnarray}
where $\mathrm{P.V.}$ denotes the Cauchy principal value,
corresponding to the singularity at $t=0$ of the integrand in
Eq.~(\ref{ei1}), and Eq.~(\ref{ei3}) is an asymptotic expansion for
$|z| \to \infty$.


\begin{thebibliography}{99}
\bibitem{feynman}
R. P. Feynman and F. L. Vernon, Ann. Phys. (N.Y.) \textbf{24}, 118 (1963).
\bibitem{leggett}
A. O. Caldeira and A. J. Leggett, Physica A \textbf{121}, 587 (1983).
\bibitem{grabert}
H. Grabert, P. Schramm and G. L. Ingold, Phys. Rep. \textbf{168}, 115 (1988).
\bibitem{caldeira}
A. O. Caldeira, H. A. Cerdeira and R. Ramaswamy, Phys. Rev. A \textbf{40},
3438 (1989).
\bibitem{unruh}
W. G. Unruh and W. H. Zurek, Phys. Rev. D \textbf{40}, 1071 (1989).
\bibitem{romero}
L. D. Romero and J. P. Paz, Phys. Rev. A \textbf{55}, 4070 (1997).
\bibitem{anglin}
J. R. Anglin, J. P. Paz and W. H. Zurek, Phys. Rev. A \textbf{55}, 4041 (1997).
\bibitem{robertson}
H. P. Robertson, Phys. Rev. \textbf{46}, 794 (1934).
\bibitem{trifonov}
D. A. Trifonov, Eur. Phys. J. B \textbf{29}, 349 (2002).
\bibitem{kruger}
J. Kr\"{u}ger, Phys. Rev. A \textbf{46}, 5385 (1992).
\bibitem{lombardo}
F. C. Lombardo and P. I. Villar, Phys. Lett.  A {\bf 336}, 16 (2005).
\bibitem{QBM1}
B. L. Hu, J. P. Paz and Y. Zhang, Phys. Rev. D \textbf{45}, 2843 (1992).
\bibitem{HZ}
B. L. Hu and Y. Zhang, Mod. Phys. Lett. A \textbf{8} 3573 (1993); Int.
J. Mod. Phys. {\bf 10}, 4537 (1995).
\bibitem{QBM3}
B. L. Hu and A. Matacz, Phys. Rev. D \textbf{49}, 6612 (1994).
\bibitem{halliwell}
C. Anastopoulos and J. J. Halliwell, Phys. Rev. D \textbf{51}, 6870 (1995).
\bibitem{halliwell-yu}
J. J. Halliwell and T. Yu, Phys. Rev. D \textbf{53}, 2012 (1996).
\bibitem{HRV}
B. L. Hu, A. Roura and E. Verdaguer, Phys. Rev. D \textbf{73}, 044002 (2004).
\bibitem{roura}
E. Calzetta, A. Roura and E. Verdaguer, Physica A \textbf{319}, 188 (2003).
\bibitem{feix}
G. Manfedi and M. R. Feix, Phys. Rev. E \textbf{62}, 4665 (2000).
\bibitem{lin}
S. Y. Lin and B. L. Hu,
``Where is the Unruh effect? - New insights from exact solutions of
uniformly accelerated detectors'', \eprint{gr-qc/0611062}.
\bibitem{ShiHu04}
K. Shiokawa and B. L. Hu, Phys. Rev. A \textbf{70}, 062106 (2004).
\bibitem{solnHPZ}
G. W. Ford and R. F. O'Connell, Phys. Rev. D \textbf{64}, 105020 (2001).
\bibitem{naik}
A. Naik \emph{et al.}, Nature \textbf{443}, 193 (2006).
\bibitem{lahaye}  M.D. LaHaye, O. Buu, B. Camarota and K.C. Schwab,
Science 304, 74 (2004).
\bibitem{buonanno}
A. Buonanno and Y. Chen, Phys. Rev.  D {\bf 64}, 042006 (2001).
\bibitem{kimble}
H. J.~Kimble  \emph{et al.}, Phys. Rev.  D {\bf 65}, 022002 (2002).


\end{thebibliography}
\end{document}